\documentclass[aps,epsf,two column,prbl]{revtex4}
\usepackage[utf8]{inputenc}
\usepackage{amsmath}
\usepackage{amsfonts}
\usepackage{subfig}
\usepackage{bbold}
\usepackage{amssymb}
\usepackage{mathrsfs}
\usepackage[final]{graphicx}
\usepackage{graphicx, psfrag}
\usepackage{float}
\usepackage{xcolor}
\usepackage[user,titleref]{zref}
\usepackage{tikz}
\usepackage{pgfplots}
\usepackage[colorlinks=true, citecolor=blue, urlcolor = blue, linkcolor= red, bookmarks=true]{hyperref}
\usepackage{epstopdf}
\graphicspath{{images/}}
\usepackage{blindtext}
\def \beq{\begin{equation}}
\def \eeq{\end{equation}}
\def \bse{\begin{subequations}}
\def \ese{\end{subequations}}
\def \bea{\begin{eqnarray}}
\def \eea{\end{eqnarray}}
\def \bem{\begin{displaymath}}
\def \eem{\end{displaymath}}
\def \bem{\begin{bmatrix}}
\def \eem{\end{bmatrix}}

\def \nn{\nonumber}

\def \bc{\begin{center}}
\def \ec{\end{center}}

\begin{document}
\title{\textbf{Revisiting Andreev processes in superconductor-graphene-superconductor (SGS) Josephson junctions: Comparison with experimental results}}
\author{Shahrukh Salim, Rahul Marathe and Sankalpa Ghosh}
\affiliation{Department of Physics, Indian Institute of Technology, Delhi, Haus Khas, 110016, New Delhi, INDIA.}
\email{sankalpa@physics.iitd.ac.in}
\begin{abstract}
In view of the recent progress in experiments on charge transport through various Josephson junctions made out of graphene, we have made a careful comparison between the theory and some of the available experimental results. Within the framework of a transfer matrix approach, we have first analytically derived the spectrum of Andreev bound states (ABS) in a superconductor -graphene-superconductor (SGS) junction for a wide range of experimentally relevant parameters. We have particularly 
considered the case of monolayer graphene (MLG). The theoretical results can  account for  both the retro Andreev reflection (RAR) and the specular Andreev reflection (SAR) in the relevant parameter range. 
Using the ABS spectrum we have evaluated the current through such junctions and the junction conductance from the analytically derived expressions at different bias voltages for a range of other system parameters directly taken from the experimental works. These theoretical results have then been compared with experimental results. 
Evaluated current and the conductance show scaling behaviour with change in the junction length and agree well with the experimental results. In the relevant parameter regime where the SAR process is dominant, the calculated values of the current and the conductivity have been found much lower than the corresponding values observed when the RAR process is dominant. 

\end{abstract} 
\maketitle
\section{Introduction}
\zlabel{sec:intro}
A normal metallic conductor placed in contact with that of a superconductor forms  the so-called superconductor-normal-superconductor (SNS) Josephson junction (JJ), which has unusual electronic properties due to the formation of bound states created at excitation energies within the superconducting gap, known as Andreev bound states (ABS) \cite{Andreev}. The existence of such Andreev bound states 
explains the microscopic origin of Josephson tunneling through such junctions  and has been extensively studied for SNS JJs \cite{josephson, tinkham, BTK, Furusaki}. In the associated transport process when an electron(hole) with energy less than the superconducting gap is incident on the normal (metal) superconductor (NS) interface, it is reflected back as a hole(electron) retracing the path of 
the electron(hole). This is known as the retro Andreev reflection (RAR). During this process, a charge of $2e$ is transferred across the junction giving rise to a Josephson current. 

In a seminal work, C. W. Beenakker  \cite{beenk_sar} extended this idea to a superconductor-graphene (SG) interface. 
It was pointed out that when the Fermi energy of mono-layer graphene (MLG)  $E_{F} \gg \Delta$, where $\Delta$ is the superconducting gap, then the Andreev reflection (AR) in such junctions is similar to the commonplace RAR in SNS junctions where also the Fermi energy $E_{F} \gg \Delta$. Such RAR in SGS junction is of intra-band nature. It turns out to be that the gapless ultra-relativistic nature of  the dispersion of monolayer graphene can lead to the condition where $E_{F}$  in MLG can  be  lowered below the superconducting gap $\Delta$ across the charge neutral Dirac point through the application of suitable gate voltage.  
However, in such a situation the RAR process is suppressed, and subsequently for $E_{F} <  \Delta$, the nature of the AR from the superconductors-graphene (SG) interface
can be completely changed from intra-band to inter-band processes,  leading to specular Andreev reflection (SAR),  a phenomenon that cannot be observed in a  prototype SNS junction. Subsequent work \cite{beenk_bal} studied the maximal supercurrent through such SGS Josephson junction in the ballistic regime and how they scale with the junctions aspect ratio. The oscillatory behavior of the tunneling conductance of a Graphene-Superconductor (GS) junction with an insulating potential barrier at the interface was also theoretically investigated \cite{Shubhro}. In  another work work \cite{Linder1} the behaviour of the supercurrent in superconductor-ferromagnet-superconductor (SFS) graphene junction was also investigated where an exchange splitting was introduced in the intermediate graphene region. 
Heat transport of Dirac fermions in GS junction \cite{Linder2}, as  well as tunable supercurrent at charge neutrality point in a strained graphene-superconductor junction \cite{Linder3} were also studied. Later work \cite{Salehi} also studied thermal transport properties in SFS junction. In a graphene ferromagnet-superconductor-ferromagnet junction (FSF),  the controlling of induced spin-triplet correlation in the superconducting region by changing the Fermi energy 
on graphene region and its possible application as a spin-valve were also studied \cite{AlidoustST}.
Graphene based superconducting  junction where the superconducting region shows $d$-wave symmetry and its effect on thermal Dirac fermions was also explored  in another work \cite{Alidoust}.
\begin{figure*}
\center
\subfloat[]{\includegraphics[width=.9\columnwidth]
{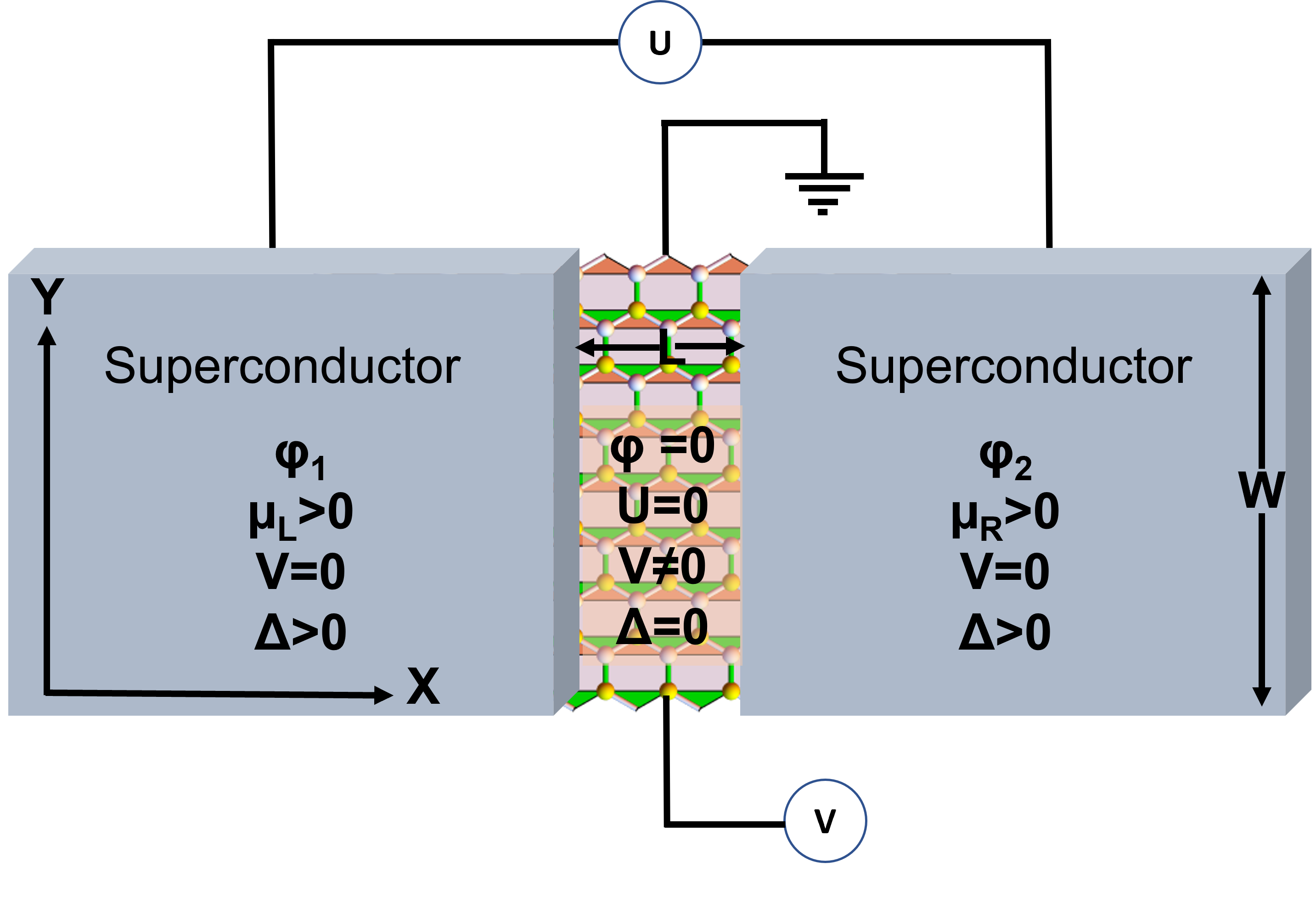}}
\subfloat[]{\includegraphics[width= .9\columnwidth]
{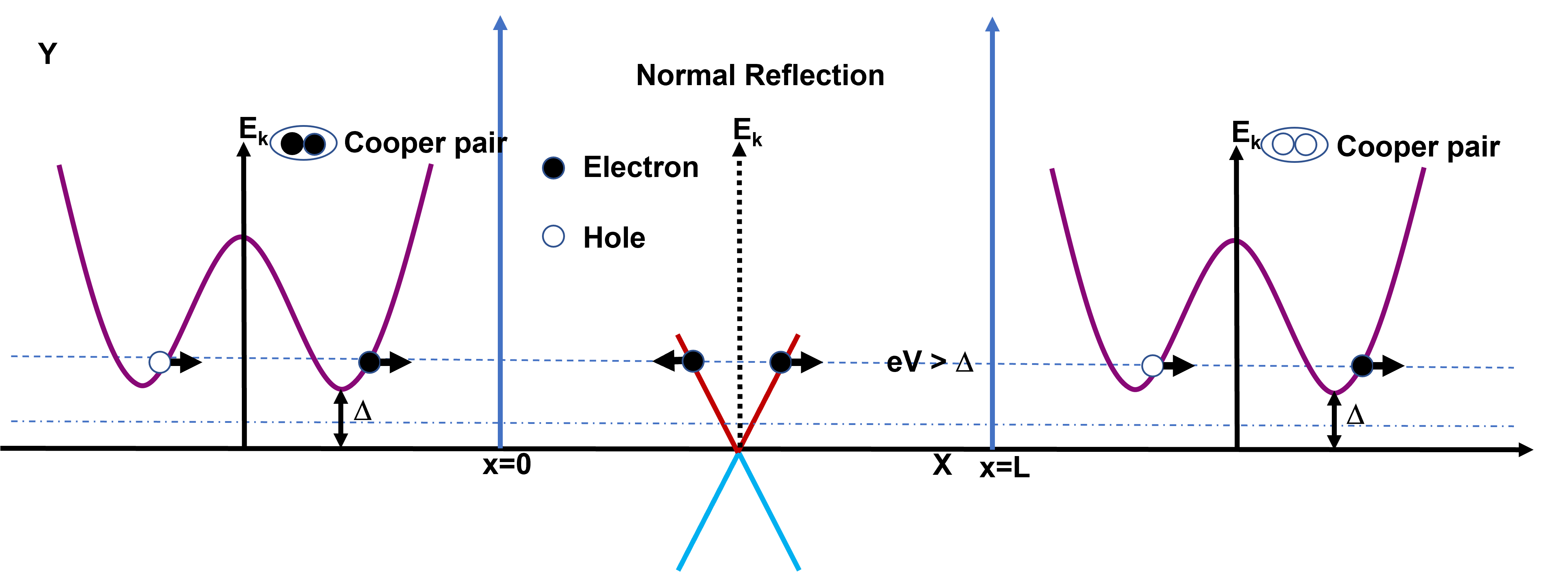}}\\
\subfloat[]{\includegraphics[width= .9\columnwidth]
{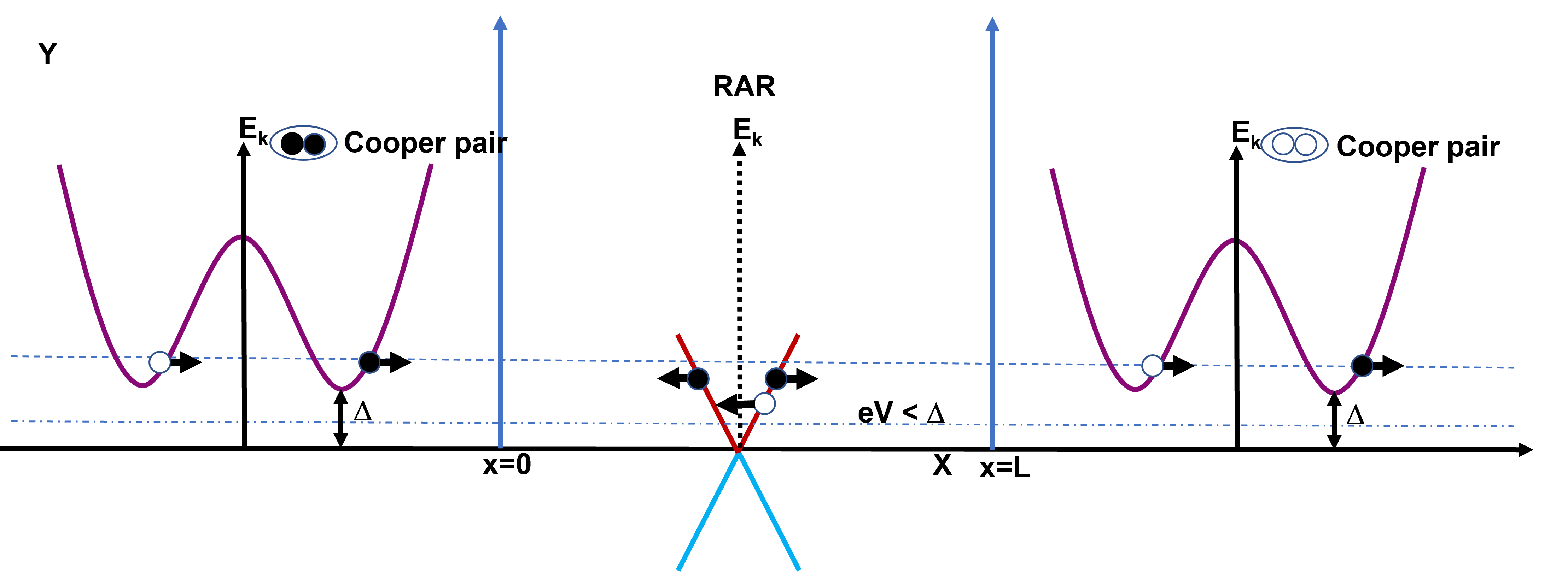}}
\subfloat[]{\includegraphics[width= .9\columnwidth]
{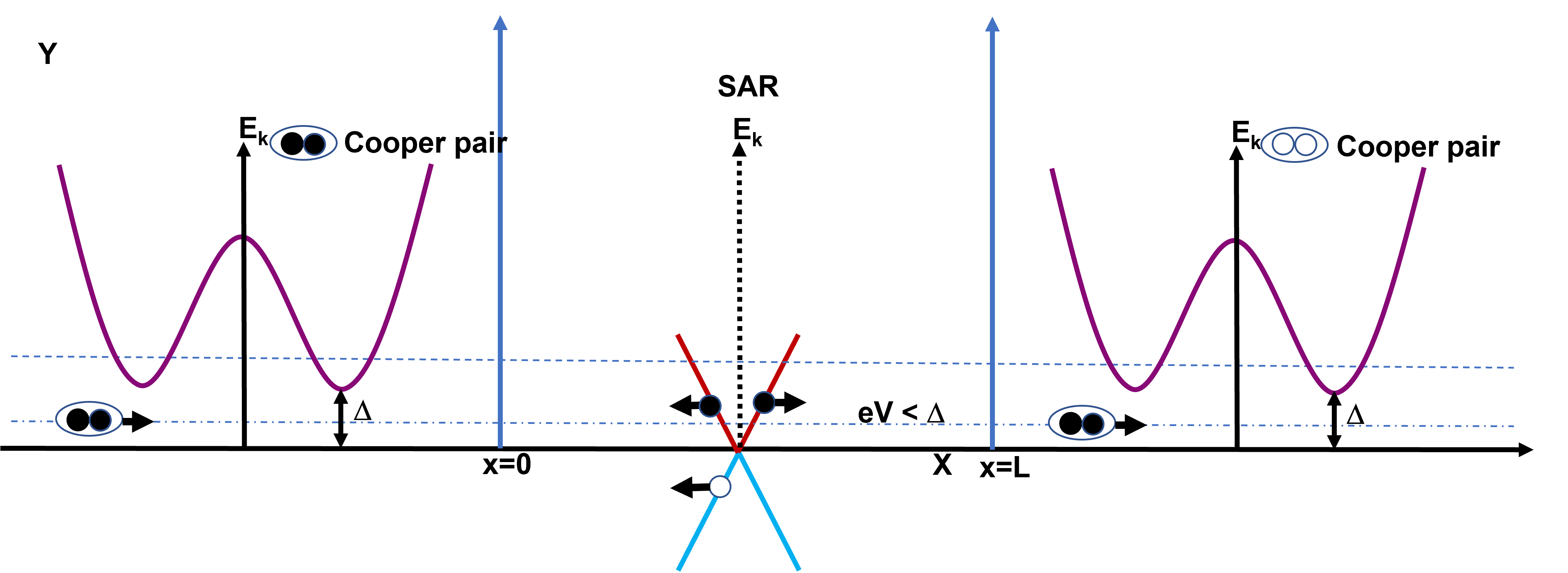}}
\caption{\label{fig:sgs}(a) Depiction of a graphene superconductor Josephson junction. The middle region corresponds to graphene and the left and right sides are made of superconductors using superconducting electrodes. $U_{0}$ is the biasing potential applied across the superconducting regions and $V$ is the gate potential applied across the graphene region. (b-d)These figures  depict various  reflection processes happening at the SGS interface. The red (upper) part of the cone is the conduction band and the blue (lower) part is the valence band in all these figures. A solid black circle represents an electron while an open circle represents a hole.
(b) Normal reflection: when the energy of the incident electron is greater than the superconducting gap. (c-d) Andreev reflection: when the energy of the incident electron is within the superconducting gap.
(c) Retro Andreev reflection (RAR) is intra  band reflection when the reflected hole lies in the same band as the incident electron. (d) Specular Andreev reflection (SAR) is inter-band reflection when the reflected hole lies in a different band.  Various quantities appeared in the figure are explained in the text.}
\end{figure*}

These remarkable theoretical predictions led to several remarkable experiments \cite{Herrero, Andrei, Popinciuc, Naomi, Dean, Bretheau, borzenets}. These were performed to understand the nature of Andreev processes and the Josephson effect  in such SGS junctions. 

Incidentally the strong fluctuation of Fermi energy at the charge neutral Dirac point  (roughly of the order of $50$~meV) in MLG due to the unavoidable presence of electron-hole puddle \cite{Yacoby}, makes it extremely challenging to lower the Fermi energy of MLG substantially below the typical superconducting gap $\Delta$. 
Consequently, most of these experiments in SGS Josephson junctions made of monolayer graphene on a given substrate thus far have been unable to observe SAR \cite{Herrero, borzenets}. 
Recently it became possible to observe the specular inter-band Andreev reflection in Van der Waals heterostructure made of bilayer graphene (BLG) - hexagonal boron nitride (hBN)  surfaces \cite{Dean} where the Fermi energy broadening near the Dirac point is relatively suppressed \cite{efetov, Soori}. A subsequent experiment that probed the nature of Andreev reflection in SGS \cite{anindya} made of Van der Waals heterostructure and MLG on hBN, was able to see the change in the nature of AR  and the suppression of RAR 
as the Fermi level is lowered through the Dirac point, but they still could not directly observe the SAR because of strong Fermi energy fluctuation at the charge neutrality point. 
In another set of experiments \cite{Andrei, Popinciuc, Naomi}, the nature of electron transport in SGS constructed with MLG that leads to such Andreev processes was investigated in detail. These experiments showed a steady improvement towards the creation of a ballistic SGS JJ.
In yet another recent experiment the spectral properties of the Andreev bound states in such systems as a function of the Fermi energy of the ML were probed in detail through tunnelling spectroscopy \cite{Bretheau}. By measuring the differential conductance through such junction as a function of energy and the superconducting phase difference in such SGS junction, the role of such bound states in carrying the Josephson current was explored in detail. The behavior of the critical Josephson current in the ballistic regime for SGS junction made of MLG on hBN substrate as a function of junction length and their scaling properties were experimentally probed in another significant experiment \cite{borzenets} that provides the details of the behavior of critical current, transmission coefficient and the normalized conductances for such junctions. Transport properties related to the Josephson effect in superconductor-graphene-normal metal (SGN) junctions were also experimentally investigated \cite{SGN}. 

In spite of these significant advancements in experimental studies and the availability of a large amount of experimental results, a careful comparison of the recent experimental results, particularly the one performed in ballistic SGS JJ, with the existing theoretical models based on the work of the \cite{BTK, beenk_sar, beenk_bal, Shubhro} is missing. Such a systematic comparison with the experimental results has the potential to engineer better microscopic theory as well as lead to better device creation. The current manuscript deals with a careful re-examination of the theoretical modelling of such junction by comparing the theoretical results with the experiments performed on ballistic SGS junctions. We have used  the well known transfer matrix framework 
to provide a unified and  general description of the AR processes in a SGS junction prepared out of monolayer graphene. Analytical expression determining the Andreev bound states for a wide range of experimental parameters is derived. After showing the resulting spectrum of such states graphically, critical Josephson current through such junctions is evaluated for the parameters range given in the experiments so that the results can be directly compared with the same. Similar calculations are also done for the transmission strength and conductance and comparisons with the experimentally observed results are made. 
A comparison is made between the theoretically calculated differential resistance with the experimentally obtained one for the parameter regime where retro Andreev reflection (RAR) dominates. The resonance peaks 
obtained in the theoretical calculation agrees well with the experimental results obtained for the ballistic junctions. Using the theoretically calculated value of the critical current and junction normal resistance we also show how their dependence on the gate voltage agrees well with the experimentally observed behaviour reported in \cite{Herrero}
 and reflects the bipolar nature of such junctions.
In the same theoretical framework we further consider the SAR process and clearly point out why the current experiments are not able to observe the SAR directly with SGS made out of MLG. The rest of the paper is organised in the following the manner: the section \zref{model}  of the  paper discusses the model and the method used to obtained the ABS for the two Andreev processes.Then we discuss the RAR processes in detail. We particularly show the behaviour of current and conductivity  for the RAR process in subsections \zref{rar} and \zref{rc} respectively.  Then we discuss the SAR processes in details in Section \zref{spar}. Lastly, in Section \zref{conc} we summarise the paper based on our comparison of the theory and the experiments. Some of the intermediate steps for the derivations in the main paper  are provided in the appendix \ref{app1}.

\section{Model and Method}
\zlabel{model}

We consider a SGS junction in a monolayer graphene sheet of length \emph{L} in the $x-y$ plane with the superconducting regions extending from $x=-\infty$ to $x=0$ and from $x =\emph{L}$ to $x=\infty$ for all $0\le y \le \emph{W}$. This is shown schematically in the Fig. \ref{fig:sgs}. 
The superconducting region is introduced through superconducting electrodes. 
 The SGS junction can be described by the DBdG equations
\begin{equation}
\begin{pmatrix}
    v_{F}(\vec{\sigma}\cdot\vec{p})-E_{F}^{\text{eff}}+\mu & \Delta e^{i\phi} \\
    \Delta e^{-i\phi} &  E_{F}^{\text{eff}}-\mu-v_{F}(\vec{\sigma}\cdot\vec{p})
\end{pmatrix}
\Psi
=E
\Psi. \label{eq:DBdG}
\end{equation}
Here $\mu$ is the general notation of the chemical potential in the  superconducting region and $\mu_{L,R}$ is the chemical potential for the left and the right superconducting region leading to the bias voltage $U_{0}$ across the junction 
\begin{equation}
\label{eq:bias}
U_{0} = \mu_{L}-\mu_{R}. 
\end{equation}
$V$ is the gate potential which is applied in the graphene region and is used to shift the Fermi level in the graphene region and is given by 
\begin{equation}
\label{eq:potentialV}
V(x)=V\theta(x)(1-\theta(x-L))
\end{equation}
where $\theta$ is the Heaviside step function.
As a result $E_{F}^{\text{eff}} $=$ E_{F}$-$V$; the effective Fermi energy in the graphene region of the junction, so that a relatively larger $k_{F}$ can be achieved in the superconducting region.
In one the experiments \cite{borzenets}, whose results are compared with the theoretical calculations in this manuscript, the Dirac point was approximately located at $V=4.8V$. The Fermi energy $E_{F}$  is chosen from a given range in ref. \cite{Fermienergy} and fixed at $3\text{eV}$ after taking into consideration
the graphene and the nature of the substrate. 
$\Psi$ is the four component wavefunction for electron and hole spinors,  $\Delta$ is the superconducting gap and $\phi_{1,2}$ are the superconducting phases in the left and right side such that
\begin{equation}\label{eq:gap}
\mathbf{\Delta}=\begin{cases}\Delta_{0}e^{i\phi_{1}}    \text{ if } -\infty<x<0 \\
 \Delta_{0}e^{i\phi_{2}}    \text{ if } L<x<\infty \\
 0 \text{, otherwise } 
 \end{cases}
\end{equation}
and the superconducting phase difference across the junction is $\phi= \phi_{1} - \phi_{2}$. 
The superconducting gap amplitude $\Delta_{0}$ is a temperature ($T_{0}$) dependent quantity and is given by \cite{tinkham}
\begin{equation}\label{eq:gapT}
\Delta_{0} = \Delta_{0}(0) \sqrt{1 - \left(\frac{T_{0}}{T_{c}}\right)^{2}},
\end{equation}
where $T_{c}$ is the critical temperature.

 The different wavefunctions for the left and right superconducting regions and the electron-hole wavefunctions in the graphene regions are given by  the stationary solutions of Eq. (\ref{eq:DBdG}).
In general these stationary solutions for the left  superconducting region (Eq. \ref{eq:superel} and Eq. \ref{eq:superhl}) and for the right superconducting regions (Eq. \ref{eq:rcsuperel} and Eq. \ref{eq:rcsuperhl}) can be written as follows:
\begin{equation}\label{eq:superel}
\psi_{Se}^{\pm}=e^{i q y \pm i k x \mp \kappa x}
\begin{pmatrix}
e^{i\beta}\\
\pm e^{\pm i(\pm\gamma+\beta)}\\
e^{-i\phi_{1}}\\
\pm e^{i(\gamma-\phi_{1})}
\end{pmatrix}
\end{equation}
\begin{equation}\label{eq:superhl}
\psi_{Sh}^{\pm}=e^{i q y \mp i k x \mp \kappa x}
\begin{pmatrix}
e^{-i\beta}\\
\mp e^{\mp i(\mp\gamma-\beta)}\\
e^{-i\phi_{1}}\\
\mp e^{i(-\gamma-\phi_{1})}
\end{pmatrix}
\end{equation} 
\begin{equation}\label{eq:rcsuperel}
\psi_{Se}^{\pm}=e^{i q y \pm i k x \mp \kappa x}
\begin{pmatrix}
e^{i\beta}\\
\pm e^{\pm i(\pm\gamma+\beta)}\\
e^{-i\phi_{2}}\\
\pm e^{i(\gamma-\phi_{2})}
\end{pmatrix}
\end{equation}
\begin{equation}\label{eq:rcsuperhl}
\psi_{Sh}^{\pm}=e^{i q y \mp i k x \mp \kappa x}
\begin{pmatrix}
e^{-i\beta}\\
\mp e^{\mp i(\mp\gamma-\beta)}\\
e^{-i\phi_{2}}\\
\mp e^{i(-\gamma-\phi_{2})}
\end{pmatrix}
\end{equation} 
These wavefunctions represent the electron-like and hole-like quasiparticles (corresponding to subscripts $e$ and $h$ respectively) moving in positive and negative $x$-direction in both the left and right regions, and differ only in terms of the superconducting phases $\phi_{1,2}$. Out of these four types of wavefunction in each region, only two are physically meaningful depending on how the $e^{\pm \kappa x}$ terms behave at $\pm \infty$, and are used in subsequent calculation given in Eq. \ref{SGSsolution}. In the left superconducting region $e^{ -\kappa x}$ terms will diverge as $x \rightarrow -\infty$.  Thus only the solutions with 
$e^{ \kappa x}$ terms survive and decide the physically meaningful electron and hole wave-functions.  By the same argument in the right superconducting region only the electron and hole wavefunctions with $e^{ -\kappa x}$ survive. These considerations lead to the choice of 
electron and  hole wavefunctions used in Eq. \ref{SGSsolution} and consistent with the same used in ref. \cite{beenk_sar}. 
The wavefunction for the particles in the graphene region with wave vector  $k_{x}, k_{y}$, where $k_{y}=q = \frac{2n_{y}\pi}{W}~ (\text{with}~n_{y} \in \mathcal{I}$ and energy $E$)  are  given by
\begin{equation} 
    \psi_{Ge}^{\pm} (x,y) = \frac{e^{i({\pm}k_x x+k_y y)}}{\sqrt{2\cos\alpha}}
\begin{pmatrix}
e^{\mp i\alpha/2}\\
{\pm}e^{{\pm}i\alpha/2}\\
0\\
0
\end{pmatrix}
\end{equation}
\begin{equation}
 \psi_{Gh}^{\pm}(x,y) = \frac{e^{i({\pm}k_x x+k_y y)}}{\sqrt{2\cos\alpha'}}
\begin{pmatrix}
0\\
0\\
e^{\mp i \alpha'/2}\\
{\mp}e^{{\pm}i{\alpha'/2}}
\end{pmatrix}
\end{equation} 

The total wavefunctions in the three regions are 
\bea
\Psi_{S1}& = & a\psi_{Se}^{-}+b\psi_{Sh}^{-} , -\infty<x<0 \nn \\
\Psi_{G }& = & p\psi_{Ge}^{+}+q\psi_{Ge}^{-}+r\psi_{Gh}^{+}+s\psi_{Gh}^{-} , 0<x<L\nn \\
\Psi_{S2}& = &c\psi_{Se}^{+}+d\psi_{Sh}^{+} ,L<x<\infty
\label{SGSsolution} \eea 
Here $\psi_{Se}^{-}$ represents a quasi-electron going in the -\textbf{x} direction while $\psi_{Sh}^{-}$ represents a quasi-hole going in the   +\textbf{x} direction and vice-versa.
The various parameters that appear in the above expressions for the wave-functions in Eq. (\ref{eq:superel}), Eq. (\ref{eq:superhl}),  Eq. (\ref{eq:rcsuperel}) and Eq. (\ref{eq:rcsuperhl}) are as given below 
\begin{subequations}
\begin{align}
\beta  & =  \begin{cases} \cos^{-1}(\frac{E}{\Delta_{0}}), \hspace{.5cm}\text{ if}~E<\Delta_{0} \\
 \mbox{-}i\cosh^{-1}(\frac{E}{\Delta_{0}}), \text{ if } ~E>\Delta_{0}
 \end{cases}\\
\gamma & = \sin^{-1}(\frac{\hbar v_{F}q}{|E_{F}^{\text{eff}}-U_{0}|}) \\
k_{F} & = \frac{E_{F}-V}{\hbar v_{F}}\\ 
k & =   \sqrt{\frac{(E_{F}^{\text{eff}}-U_{0})^{2}}{(\hbar v_{F})^{2}-q^{2}}} \\ 
 \kappa & =  \frac{(E_{F}^{\text{eff}}-U_{0})\Delta_{0}}{(\hbar v_{F})^{2} k}\sin\beta 
\end{align}
\label{params}
\end{subequations}
Also $k_{x}$ for electron (with subscript $_{e}$) and hole (with subscript $_{h}$) can be respectively defined as
\beq k_{e} = \frac{(E+E_{F}^{\text{eff}})}{\hbar v_{F}}\cos\alpha ; \hspace{.5cm} k_{h}=\frac{(E-E_{F}^{\text{eff}})}{\hbar v_{F}}\cos\alpha',  \label{params1}  \eeq
with  $\alpha$ and $\alpha '$ are defined as the incident angles for electron and hole with the $x$-axis respectively, 
\begin{equation}
\label{eq:incident}
\begin{aligned}
\alpha  & = &  \sin^{-1}\left( \frac{\hbar v_{F}q}{E+E_{F}^{\text{eff}}}\right) \in[-\frac{\pi}{2},\frac{\pi}{2}] \\
\alpha ' &  = &  \sin^{-1} \left( \frac{\hbar v_{F}q}{E-E_{F}^{\text{eff}}}\right) \in[-\frac{\pi}{2},\frac{\pi}{2}]
\end{aligned}
\end{equation} 
The critical angle is defined as the ratio of the two incident angles
\beq\label{eq:crit}
\alpha_{c}=\sin^{-1}\left( \frac{|E-E_{F}^{\text{eff}}|}{E+E_{F}^{\text{eff}}}\right)  
\eeq 
The boundary conditions at each junction can be written in terms of transfer matrices. To that purpose we introduce matrices $M_{1}, M_{2}, M_{3}, M_{4}$ whose explicit forms are given in the Appendix \ref{app1}.
Using these transfer matrices we can write the matching conditions at the first and second boundary respectively as 
\begin{equation}
\begin{aligned}
M_{1}\Psi_{S1} &= & M_{2}\Psi_{G} \nonumber \\
M_{3}\Psi_{G} & =& M_{4}\Psi_{S2}
\end{aligned}
\end{equation}
After some straight forward algebra we get 
\bea 
\Psi_{S2} & = & M_{4}^{-1}M_{3}M_{2}^{-1}M_{1}\Psi_{S1}\label{tm1} \\
\Psi_{S1} & = & M_{1}^{-1}M_{2}M_{3}^{-1}M_{4}\Psi_{S2}\label{tm2} \eea 
Solving the above Eqs. (\ref{tm1}) and (\ref{tm2}) we get the 
\beq\label{gendis}
|M_{4}^{-1}M_{3}M_{2}^{-1}M_{1}| = e^{\pm2i\phi} \eeq 
The above result holds for retro and specular Andreev reflection and is also valid for both long and short junctions. For numerical evaluation and comparison with experimental results we consider cases of short junctions ($\emph{L} < \xi$) as well as long junctions ($\emph{L} > \xi$). For almost all the cases considered below 
the junction width $\emph{W} > \emph{L}, \xi$. Here $\xi$ is the coherence length of the superconductor.
We shall now discuss various limiting cases in the following sections.
\subsection{Retro Andreev Reflection and the contribution to Josephson current}
\zlabel{rar}
The retro Andreev reflection (RAR) \cite{kulik,golubov}
that has been widely studied in both SNS and SGS junctions, happens when $E <  E_{F}^{eff}$. In this case both the incident particle and the reflected particle lies in the conduction band and the conditions for the same is given by 
\beq  k_{e}  =  \frac{E_{F}^{\text{eff}}}{\hbar v_{F}}\cos\alpha = -k_{h}  \label{RAR} \eeq
leading to 
\beq
\alpha '  =  -\alpha  \nonumber \eeq 

\begin{figure*}[!htb]
\subfloat[]{\includegraphics[width=.5\columnwidth]{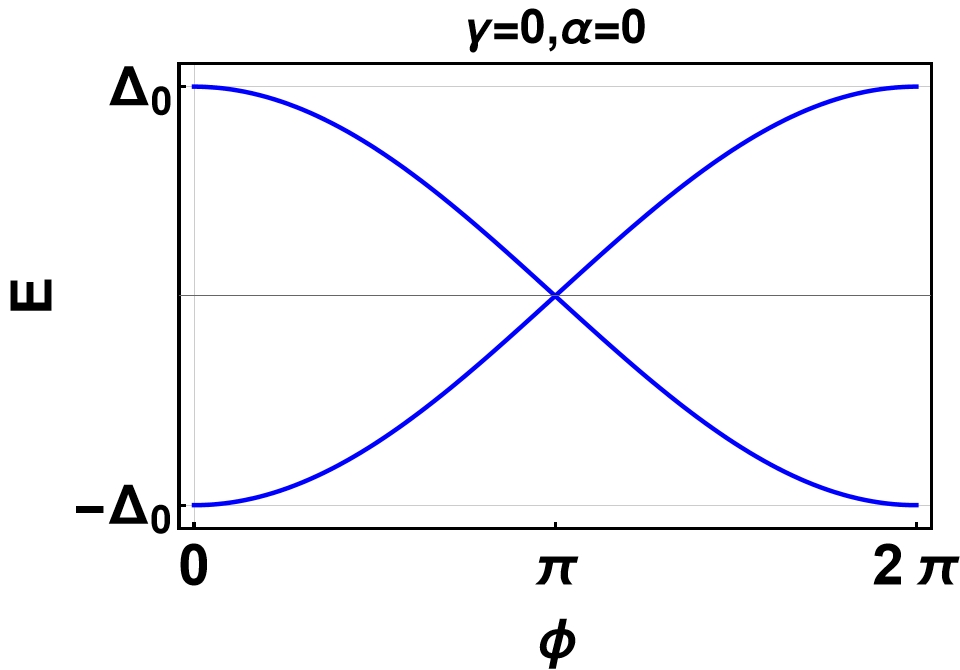}} 
\subfloat[]{\includegraphics[width=.5\columnwidth]{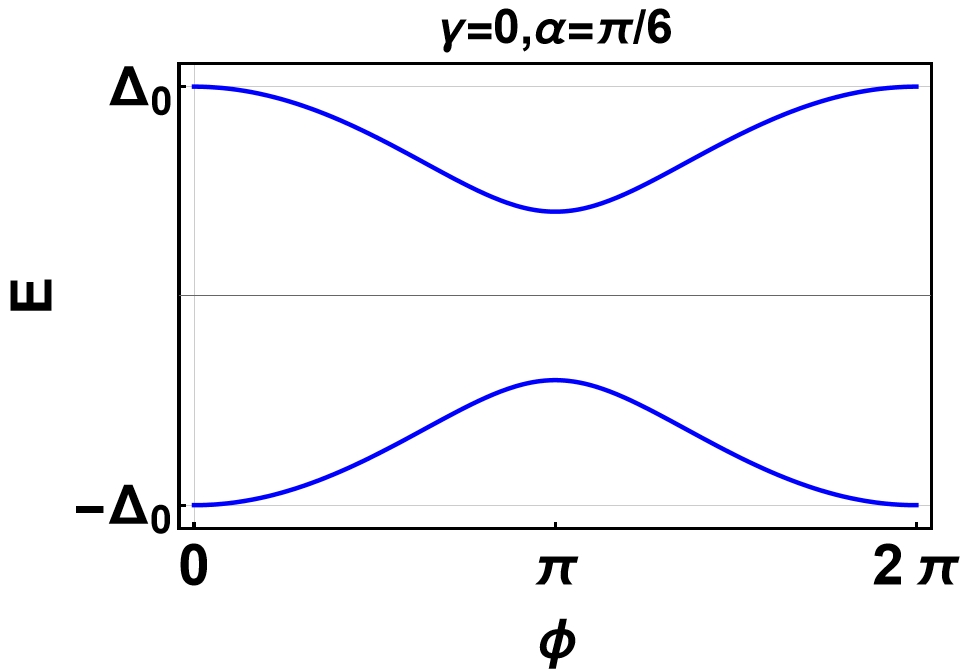}} 
\subfloat[]{\includegraphics[width=.5\columnwidth]{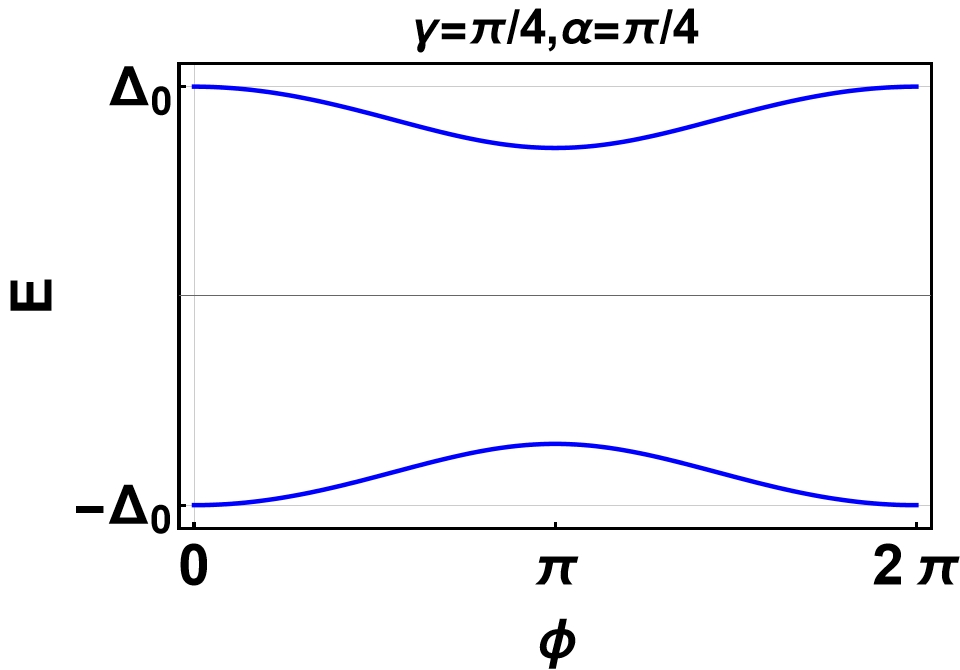}}
\subfloat[]{\includegraphics[width=.5\columnwidth]{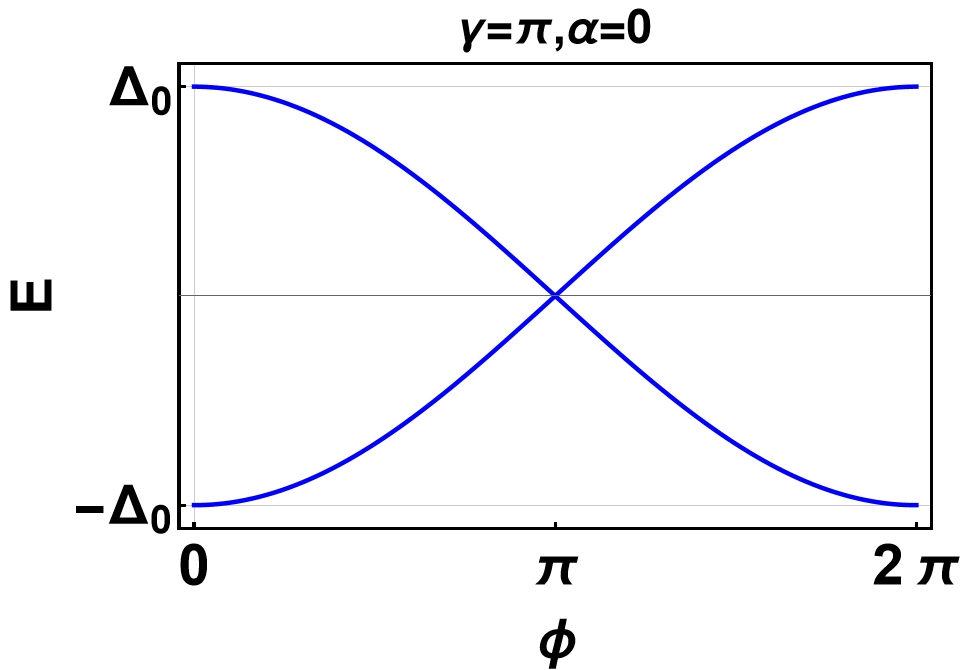}}  \\
\subfloat[]{\includegraphics[width=.52\columnwidth]{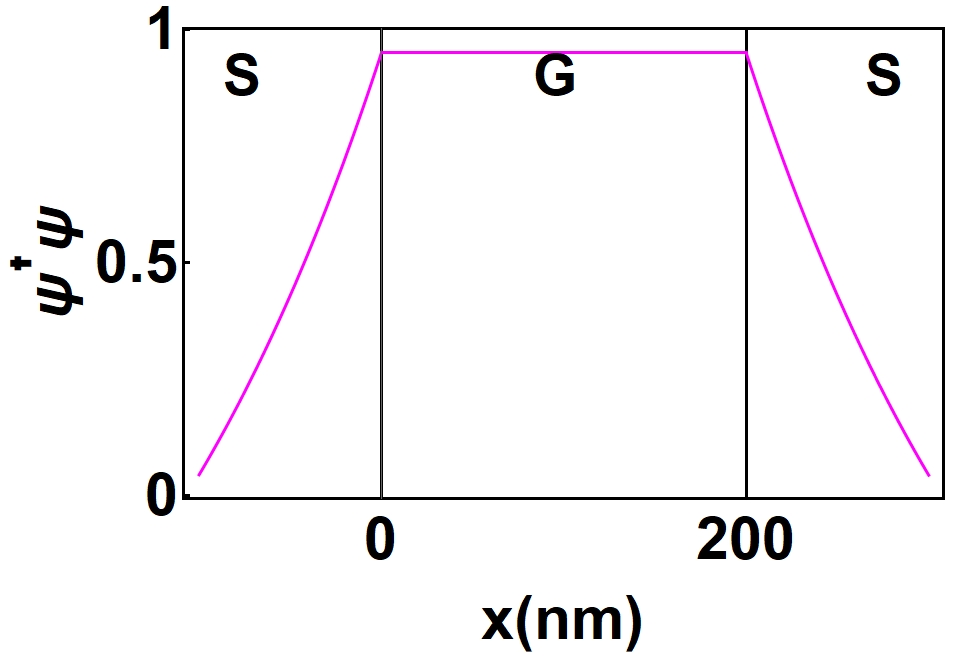}} 
\subfloat[]{\includegraphics[width=.47\columnwidth]{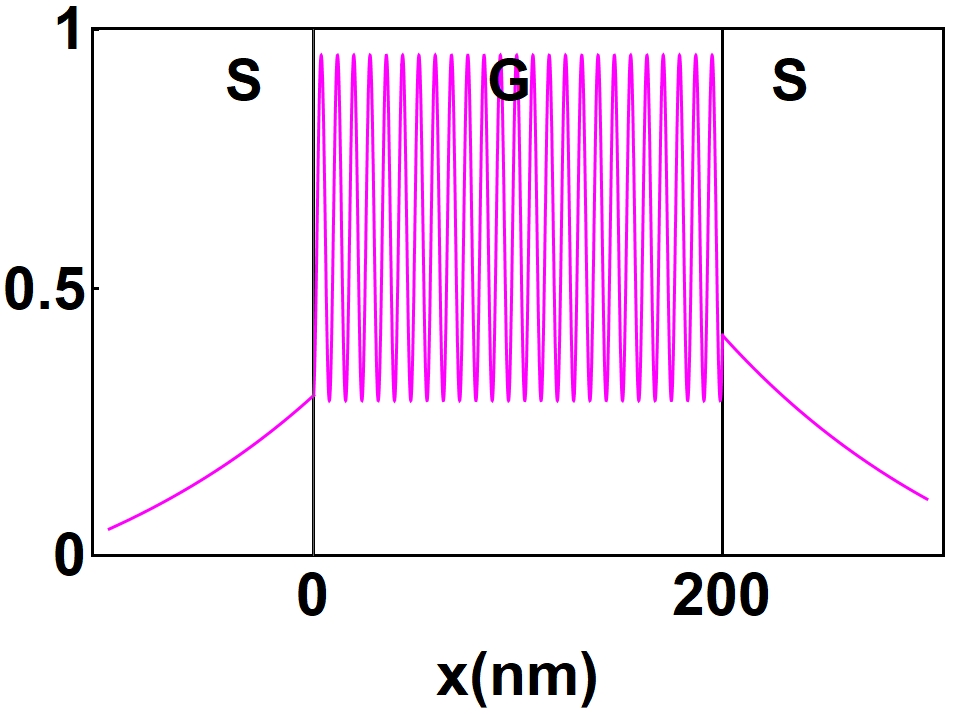}}
\subfloat[]{\includegraphics[width=.47\columnwidth]{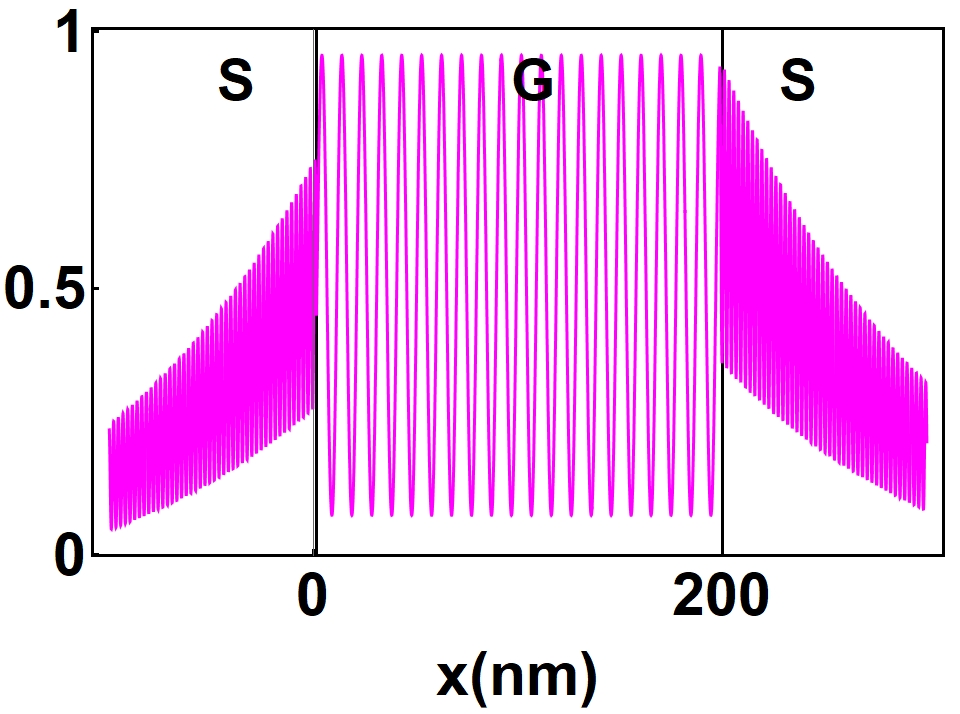}}
\subfloat[]{\includegraphics[width=.47\columnwidth]{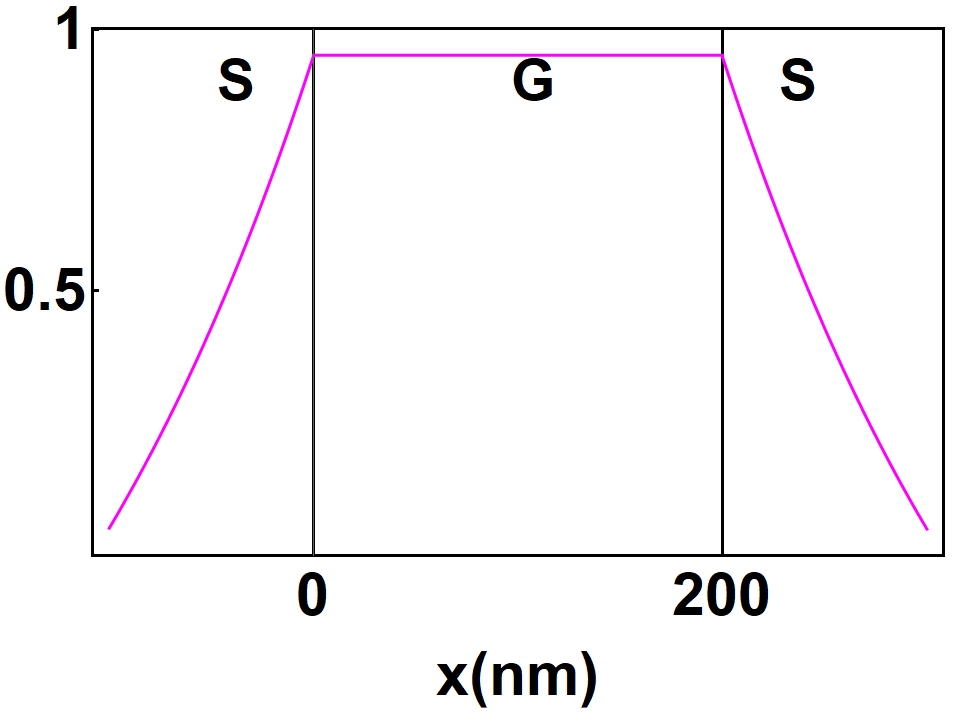}}
\caption{(a-d) Dispersion for retro reflection for different values of $\gamma$ and $\phi$. It is clear from these figures that the gap  closes for $\alpha=0,\gamma=0,\pi$. For all other values the gap opens up and we can see two distinct bands. (e-h) Probability distribution for the wave functions corresponding to the respective dispersion in the (a-d). The length $\emph{L}$ of the graphene region is 200nm for the figures, $E_{F}$ is $3.0$eV, $\Delta_{0}$ is $1.2 \text{meV} $, $\phi_{1} = \frac{\pi}{4}$ and $\phi_{2}= \frac{\pi}{6}$ for the all the figures.}
\label{retall1}
\end{figure*}

Substituting  these conditions in Eq. (\ref{gendis}) we get the conditions for the Andreev bound states (ABS) which lie within the superconducting gap, as 
\begin{widetext}
\beq \label{retgen1}
\cos ^{2}Lk_{e}\cos ^{2}\alpha (-1+2\cos 2\beta +\cos 2\gamma)  = 2 \cos ^{2}\gamma\lbrace\cos ^{2}\alpha\cos\phi +\sin ^{2}Lk_{e}(\sin ^{2}\alpha -\cos 2\beta)\rbrace \eeq
\end{widetext}
which gives the dispersion as 
\beq  E =  \pm \Delta_{0}\left[1 -t\sin ^{2}\frac{\phi}{2}\right]^{\frac{1}{2}} \label{retgen} 
\eeq 
with 
\beq t = \frac{\cos ^{2}\alpha \cos ^{2}\gamma}{\cos ^{2}Lk_{e}\cos ^{2}\alpha +\cos ^{2}\gamma\sin ^{2}Lk_{e}} \label{transmission} \eeq 
It may be noted that the dispersion given in Eq. (\ref{retgen1}) can be recast in the form of an equation of a conic section for both RAR and SAR, namely 
\beq A x^{2} + B y^{2} + 2C xy = D \label{conic}\eeq 
where $x = \cos Lk_{e}$ and $y = \sin Lk_{e}$, and, 
\begin{subequations}\label{disparam} 
\begin{align}
A  = & -2 \cos ^{2} \alpha( \sin^{2} \gamma - \cos 2\beta)  \\
B  = & - 2 \cos^{2} \gamma ( \sin^{2} \alpha + \cos 2\beta ) \\
C  = & 2 \cos \gamma \cos \alpha \sin 2 \beta  \\
D = & 2 \cos \alpha \cos^{2}\gamma \cos \phi 
\end{align}
\end{subequations}
For the case of RAR, $C=0$. 
The thin (very short) junction limit is obtained by setting $\gamma=0$, when the Fermi wavelength $\lambda_{F}$ becomes much smaller than the coherence length $\xi$ of the superconductor i.e. $\lambda_{F} \ll \xi$. 
The two bound energy states that lie within the superconducting gap $\Delta_{0}$ are still given by Eq. (\ref{retgen})
with 
\beq t (\gamma=0) = \frac{\cos ^{2}\alpha}{\cos ^{2}\alpha \cos ^{2}Lk_{e}+\sin ^{2}Lk_{e}} \label{transmissionthin} \eeq 
 The expression (\ref{retgen}) was derived in various forms \cite{krishnendu,golubov,haberkorn}.  
 It may be noted that in the thin junction limit, $t$ in (\ref{transmission}) takes exactly the form of the transmission coefficient $T$ of the barrier transmission problem for  massless relativistic fermion (quasiparticles of monolayer graphene) for a potential barrier of height $V_{0}$ and width $d$, in the limit where the energy of the quasiparticle is much less than the barrier height \cite{klen}. The 
 expression of $T$ in that case is given by 
 \beq
T = \frac{\cos ^{2}\alpha}{\cos ^{2}\alpha \cos ^{2}k'_{x}d + \sin ^{2}k'_{x}d} \nonumber 
\eeq
where $\alpha$ is again the incident angle of such quasiparticle on this barrier, and $k'_{x}d = -2\pi l\sqrt{1 - 2\varepsilon + \varepsilon^{2}\cos ^{2}\alpha}$, and $l$ is the dimensionless barrier width given by $l \equiv V_{0}d/(2\pi\hbar v_{F})$ and the dimensionless energy $\varepsilon \equiv E/V_{0}$.  
This allows us to understand the similarity in the barrier transmission problem and 
the transport through SGS junction. 
Accordingly we shall call $t$ in the present problem as the transmission coefficient.  
The bound state energy as a function of $\phi$, the superconducting phase difference, for different values of $\gamma$ and $\alpha$ are plotted in the Fig. \ref{retall1}(a)-(d).  Fig. \ref{retall1}(e)-(h) show a representative 
probability density ($|\psi|^{2}$) for such bound state in each of the corresponding cases. 
\\~\\
Further, there is also a crossing between the Andreev bound states  at $\phi= \pi$, whenever the incident angle $\alpha =0$, and $\gamma = 0~ \text{or}~\pi$ and the dispersion takes a simpler form, namely 
\beq E_{\pm}=\pm \Delta_{0} \cos\frac{\phi}{2} \label{Bernouli} \eeq 
For every other value of $\alpha$ and $\gamma$ there is always a gap between the two states. The ABS are responsible for the Josephson current and we shall evaluate them in the next section and compare them with the experimental results. 

\begin{figure}[!htb]
\includegraphics[width = \columnwidth]{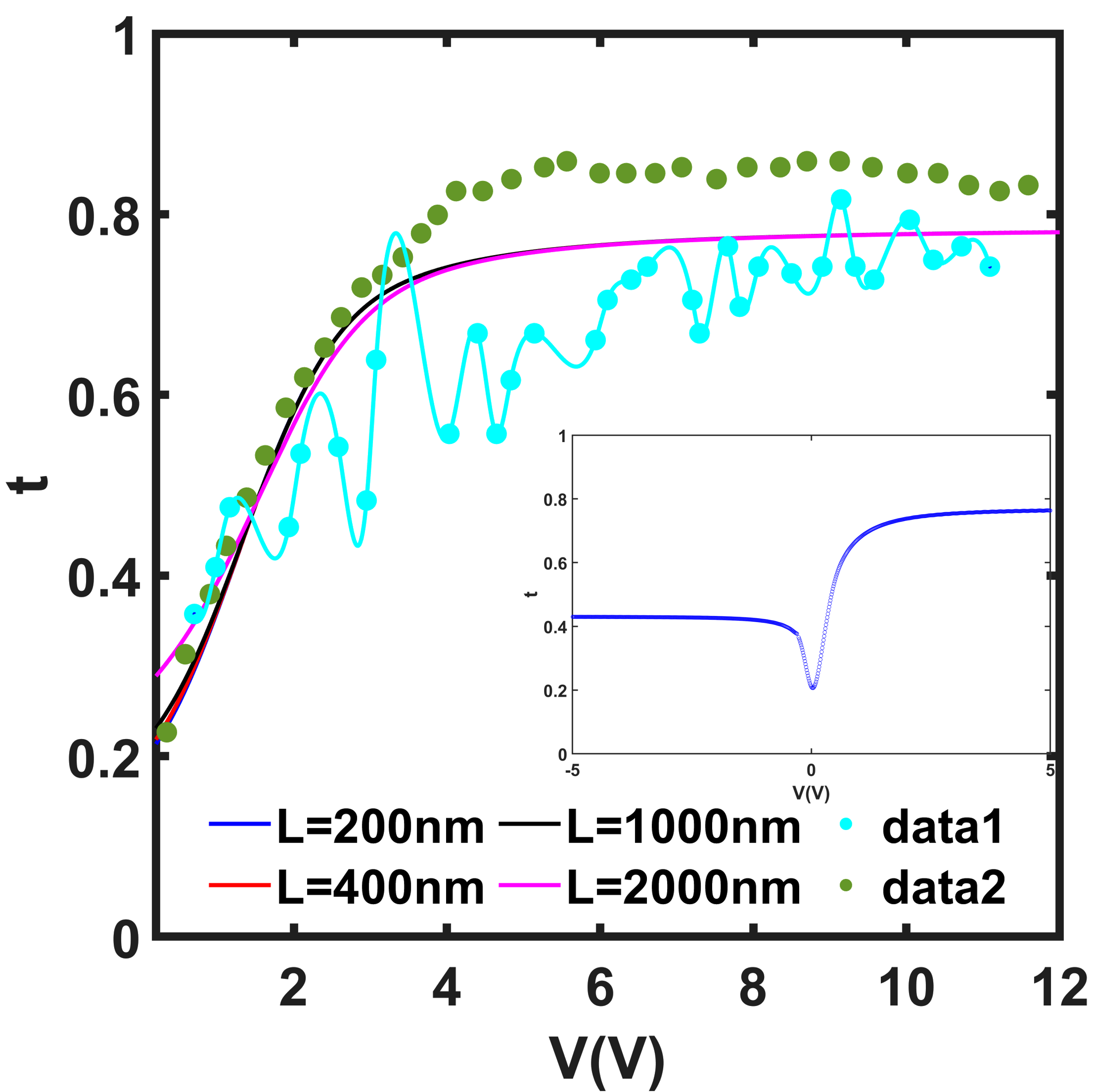}
\caption{Variation of the transmission for different values of gate voltage which is being measured w.r.t the Dirac point.  The experimental data has been taken from the paper by  Borzenets et al. \cite{borzenets} the length of the junction is 200nm. The scattered points in cyan are connected with a continuous line to make oscillations in the experiment clearly visible. It may be noted that the two sets of scattered points in the experimental work  in ref. \cite{borzenets} are obtained through two different methods. The inset shows the behaviour of the transmission $t$ defined in Eq. (\ref{transmission}) as a function of the gate voltage on both sides of the Dirac point. The transmission is asymmetric around the Dirac point. The Dirac point has been shifted to $0$ V in accordance with ref. \cite{borzenets}.}
\label{fig:ret_tr}
\end{figure}

In order to calculate the current across the junction we  need to differentiate the  energy dispersion relation Eq. (\ref{retgen}) for the subgap ABS  with respect to the superconducting phase difference $\phi=\phi_{1}-\phi_{2}$. The Josephson current $I$ \cite{kulik,krishnendu,BTK,golubov,haberkorn} across the junction at some temperature $T_{0}$ is given by the 
\begin{equation}\label{jcur1}
I(\phi,V,U_{0},T_{0})=\frac{4e}{\hbar}\sum_{n=\pm} \sum_{q=-k_{F}} ^{k_{F}} \left(\frac{\partial E}{\partial\phi}\right)_{(E=E_{n})}\tilde f(E_{n})
 \end{equation}
where$\tilde{f}(E_{n}) = f_{1}(E_{n}-\mu_{L}) -f_{2}(E_{n}-\mu_{R}) $ and $f(E_{\pm}) = \frac{1}{\exp(\frac{E_{\pm}}{k_{B}T_{0}})+1}$ is the Fermi distribution function at some temperature $T_{0}$ (the reference point of the energy is taken as the Fermi energy itself), $k_{B}$ is the Boltzmann constant. After substituting the relevant quantities in the above equation the current now reads
\begin{widetext}
\begin{equation}
I(\phi,V,U_{0},T_{0})=I_{0}\int^{\frac{\pi}{2}} _{-\frac{\pi}{2}}d\gamma \frac{t \cos\gamma \sin\phi}{\sqrt{1-t\
sin ^{2}\frac{\phi}{2}}} 
\tanh\left(\frac{\Delta_{0}}{2k_{B}T_{0}}\sqrt{1-t\sin ^{2}\frac{\phi}{2}}- \frac{eU_{0}}{2k_{B}T_{0}}\right)\label{JC1} 
\end{equation}
\end{widetext}
\text{where}
\beq I_{0}=\frac{e\Delta_{0}}{\hbar} \frac{E_{F}^{\text{eff}}}{2 \pi \hbar v_{F}} \cdot \emph{L} \label{universal} \eeq
is the bias independent current defined in \cite{beenk_bal, UCF} upto a constant factor. It may be noted 
that the first term in the expression of $I_{0}$ is the universal limit of the critical current fluctuations for a mesoscopic Josephson junction \cite{UCF}. The subsequent terms shows a linear scaling with the junction length $\emph{L}$ and how the gate voltage $V$ gives a bipolar behaviour around the Dirac point. 
This also clearly shows that at the Dirac point, where $E_{F}^{\text{eff}}=0$, this current will also go to $0$. 
It may  be pointed out that the expression (\ref{JC1}) is more general as compared to the one obtained in \cite{beenk_bal}. In the subsequent discussion we shall show that this expression agrees well 
with the experimental results obtained for a ballistic junctions over a wide range of experimental parameters.
It may be noted that the summation in the  expression (\ref{jcur1}) over the wave-vector $q$ is converted into an integration in the expression (\ref{JC1}). This integration can be performed either by integrating over the angle $\gamma$ which is a $q$ dependent quantity and defined in Eqs. (\ref{params}).
Alternatively one can integrate over $q$ and this is implemented while evaluating the expression (\ref{conduct}). The transmission $t$ defined in Eq. (\ref{transmission}) is an asymmetric function of gate voltage $V$ as shown in the inset of Fig. \ref{fig:ret_tr}. This behaviour impacts the behaviour of
the critical current in (\ref{JC1}) and the normal resistance in (\ref{conduct}).
\begin{figure}[!htb]
\includegraphics[width =\columnwidth]{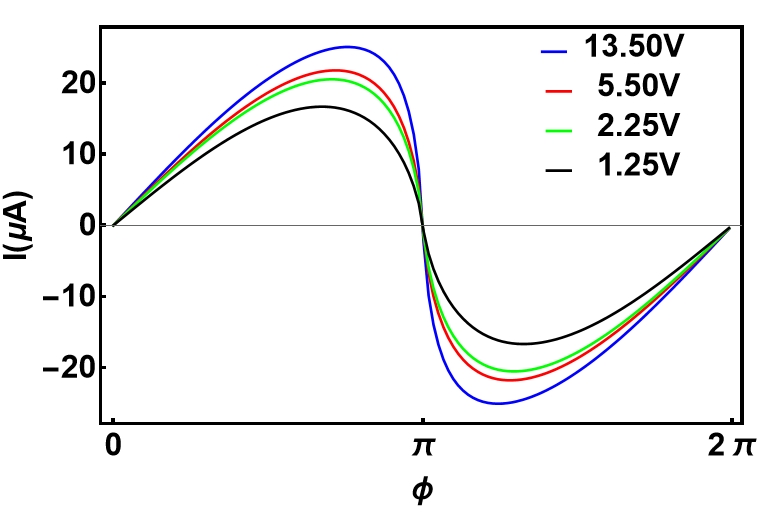}
\caption{Variation of the Josephson current with the superconducting phase difference for different values of gate voltage. A sinusoidal relation is obtained. The Josephson current is maximum for the largest value of the gate voltage.}
\label{JCFig}
\end{figure}

\begin{figure*}[!htb]
\centering
\subfloat[]{\includegraphics[width=0.45\columnwidth]{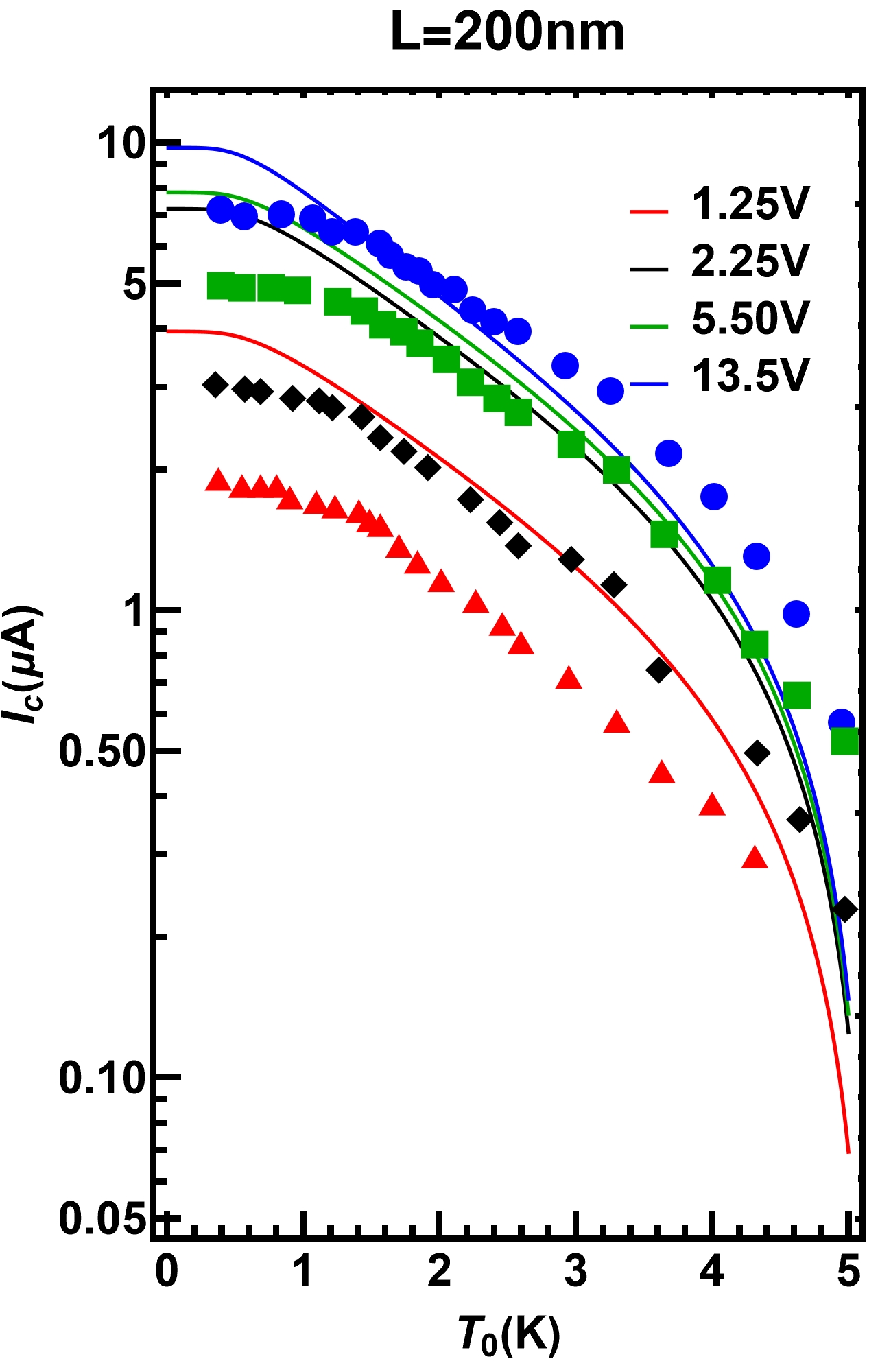}} 
\subfloat[]{\includegraphics[width=0.45\columnwidth]{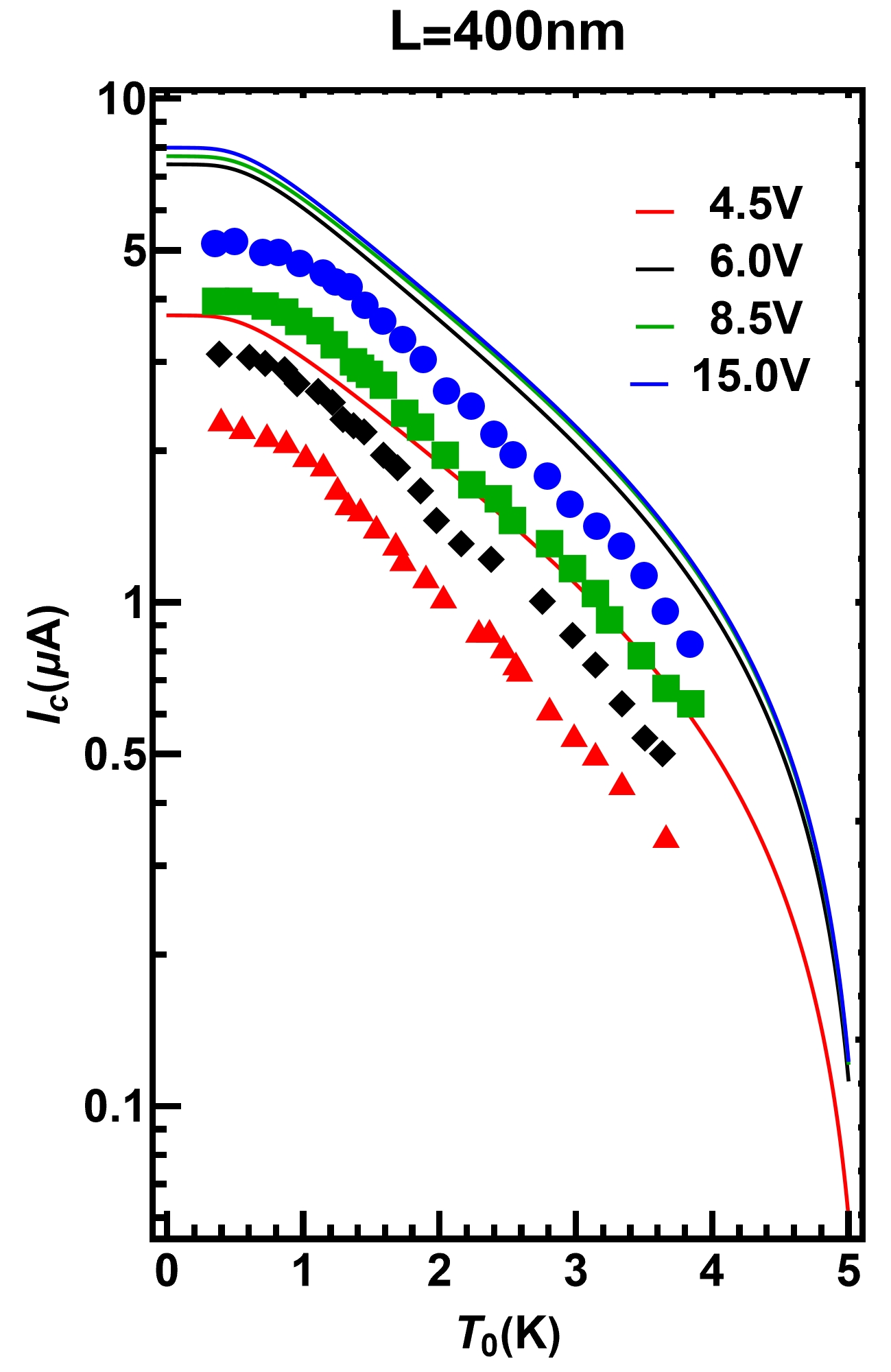}}
\subfloat[]{\includegraphics[width=0.45\columnwidth]{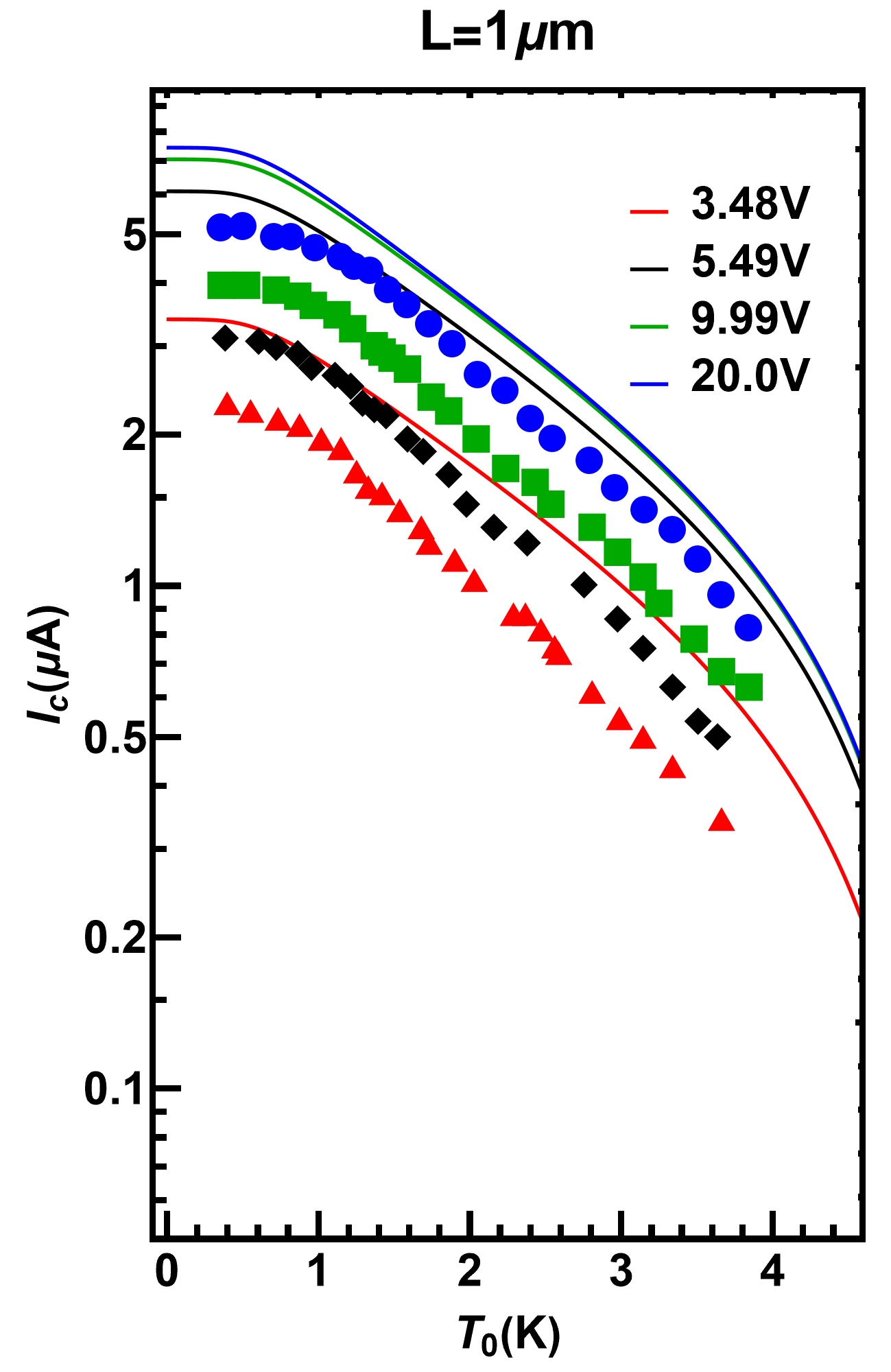}}
\subfloat[]{\includegraphics[width=0.45\columnwidth]{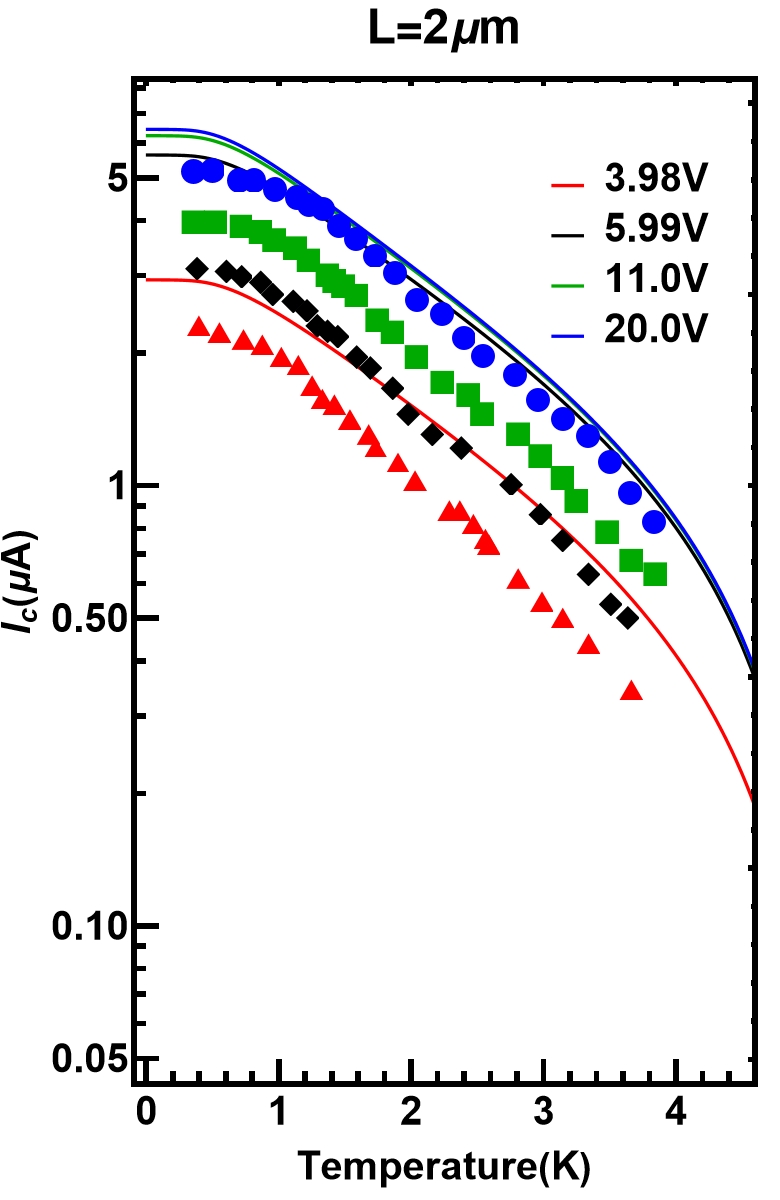}} 
\caption{Variation of the Josephson current on a semi-log scale with the temperature for different values of the gate voltage for different junction lengths $\emph{L}$. The values of $\emph{V}$ are shown relative to the Dirac point $\sim 4.8V$. The continuous lines  correspond to theoretical plots whereas the scattered symbols with the same  colour correspond to the experimental data for a given gate voltage which is   taken from the Fig. 2 of the paper by  Borzenets et al.\cite{borzenets}. The first panel (a) corresponds to a short junction, the figure (b) is in the vicinity of the coherence length but is still short. The last panels (c-d) corresponds to long junctions.}
\label{JJexpt} 
\end{figure*}

\begin{figure*}[!htb]
\subfloat[]{\includegraphics[width=.33\textwidth]{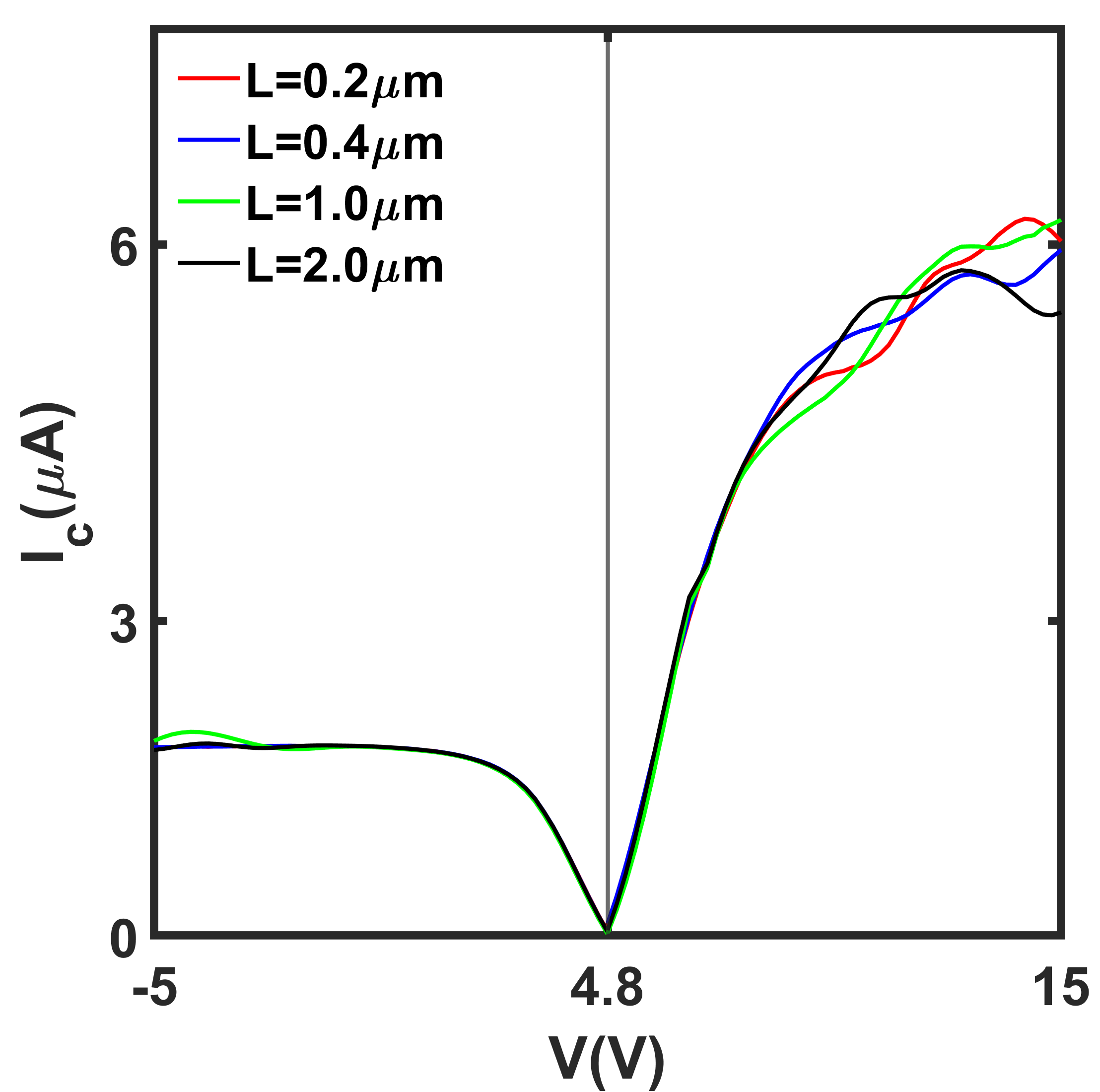}}
\subfloat[]{\includegraphics[width=.33\textwidth]{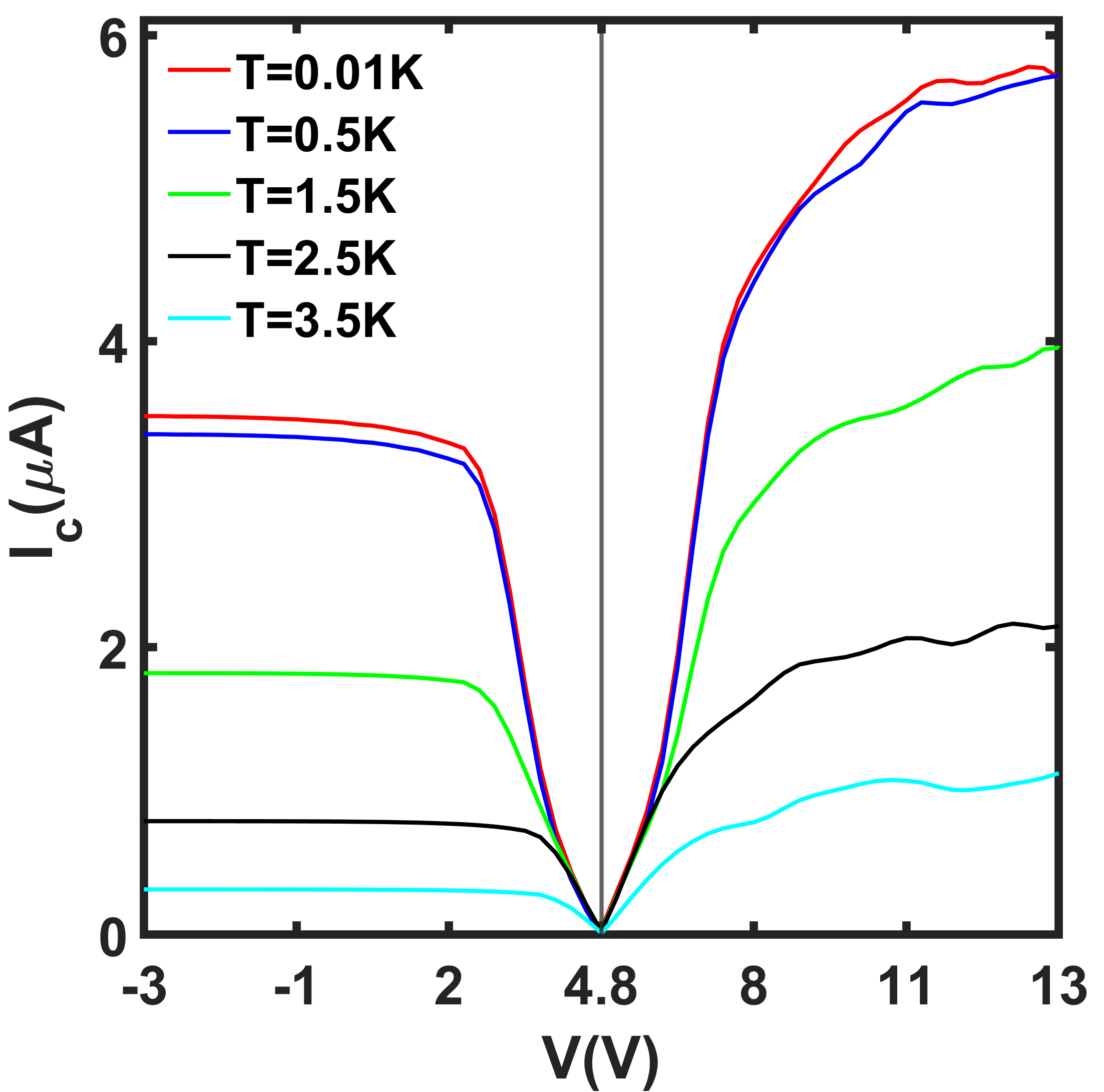}}
\subfloat[]{\includegraphics[width=.33\textwidth]{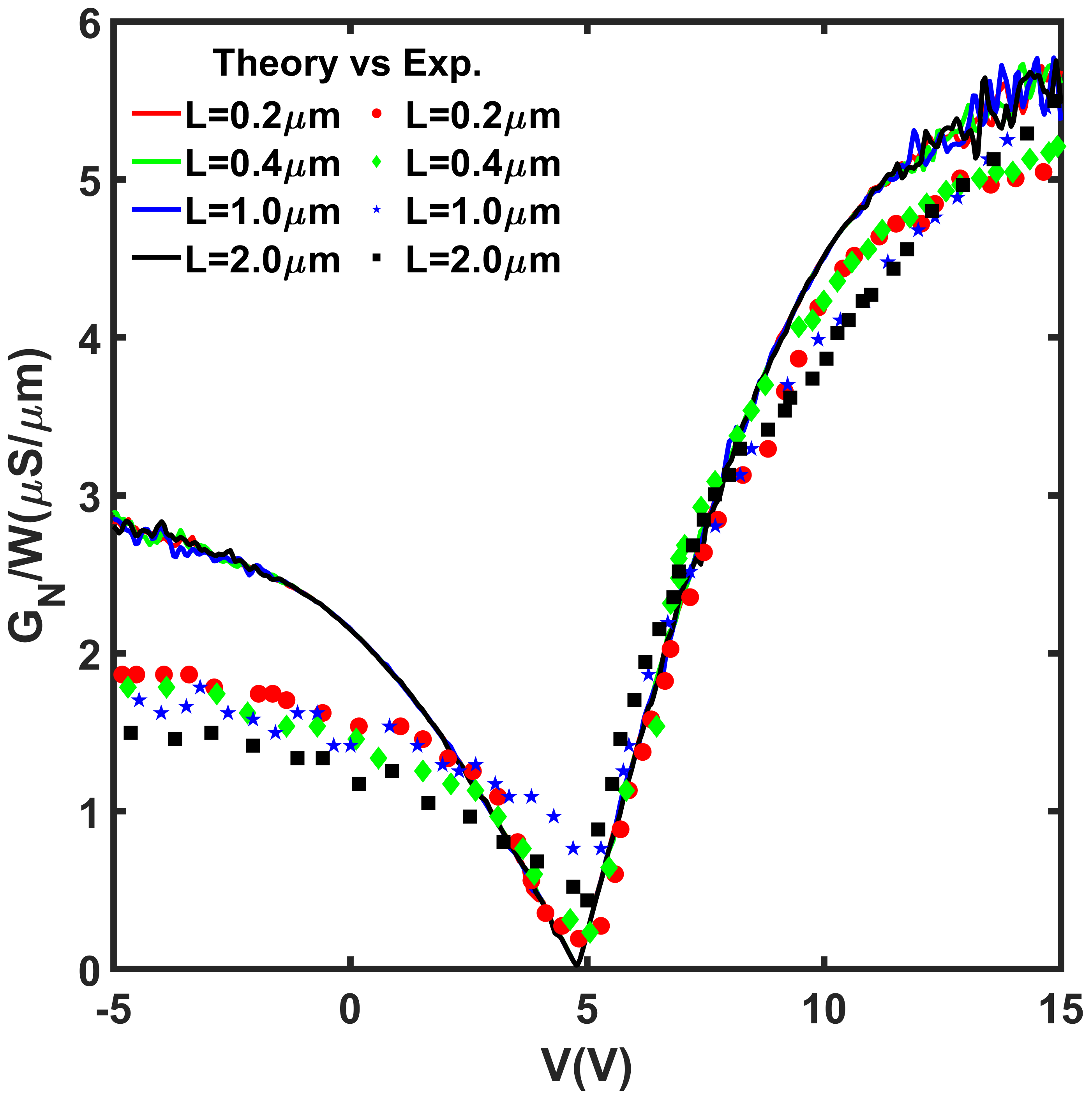}}
\caption{(a) Critical Current $I_{c}$ vs. gate voltage $V$  plot for various junction lengths (shown in the figure legend) for a fixed temperature of 2.5K.
The critical temperature $T_{c}=5.5K$ and the coherence length $\xi=550 nm$ are chosen following experimental reference \cite{borzenets}.  (b) Critical Current $I_{c}$ vs. gate voltage $V$   for a fixed junction length $L=200nm$ for increasing temperature. Temperatures are again shown in the legend.
(c) Variation of the  conductance normalised by the width ($\emph{W}$) of the junction for different values of gate voltage. Here $G_{N}$ stands for the normal state conductance and is calculated using Eq. (\ref{conduct}). The different curves correspond to the different values of the length $L$ of the barrier.   $\Delta_{0}(T_{0}=0)$ is $1.2~meV$ for all the curves corresponding to various lengths of the junction.
The scattered points are extracted from the plot  in Fig. 3(a) of the experimental paper \cite{borzenets}(after shifting the Dirac point at the appropriate $V$) whereas the continuous lines represent the theoretical calculation.}
\label{fig:ret_c2}
\end{figure*}

In Fig. \ref{JCFig} the current $I$ in Eq. (\ref{JC1}), is plotted as a function of $\phi$ for a fixed $(\emph{L, W})$ for different values of the gate voltage. 
The current $I$ becomes zero when the phase difference between the superconductors is $\pi$, which is a common feature for such kind of junctions. As the gate voltage is increased the peak of the Josephson current shifts towards the right. 
The maximum value of the Josephson current $I$ is called the critical Josephson current $I_{c}$. 
The variations of $I_{c}$ with different quantities are plotted in Fig. \ref{JJexpt} and Fig. \ref{fig:ret_c2}(a) and will be discussed in the later part of this section. 
Please note that unless otherwise states the superconducting gap $\Delta_{0}$ is always taken at the absolute zero temperature $T_{0}=0$. 
It may be noted that in the Eq. (\ref{JC1}), $t = t(\gamma)$ and we express the ABS energy in terms of $t$ using (\ref{retgen}). 
The current flowing in the junction is a function of the phase difference between the two superconductors as the energy of these subgap ABS is a function of this phase difference \cite{golubov,haberkorn} as seen in Fig. \ref{retall1}(a-d). On setting the bias potential $ U_{0}=0$ the above equation (\ref{JC1}) will now read;

\begin{widetext}
\begin{align}\label{JC2}
 I(\phi,V,T_{0})= I_{0}\int^{\frac{\pi}{2}} _{\frac{-\pi}{2}}d\gamma \frac{t \cos\gamma \sin\phi}{\sqrt{1-t\
sin ^{2}\frac{\phi}{2}}} \tanh\left(\frac{\Delta_{0}}{2k_{B}T_{0}}\sqrt{1-t\sin ^{2}\frac{\phi}{2}}\right)
\end{align}
\end{widetext}

The transmission coefficient $t$ is plotted as a function of gate voltage and compared with the experimental results 
\cite{borzenets} in Fig. \ref{fig:ret_tr}. The oscillation of the transmission coefficient and its eventual saturation with the increasing gate voltage is visible in the both the experimental and the theoretical data. There is a special case when the criteria for RAR is violated and we get SAR which will be discussed in section \zref{spar}. We will now try to compare the results with the existing experimental literature.
 
To compare these results we have plotted the critical Josephson current through such junctions in Fig \ref{JJexpt} using this expression (\ref{JC1}) along 
with the experimental data of Borzenets {\it et al.} \cite{borzenets} where they have  studied ballistic graphene Josephson junctions with Molybdenum-Rhenium super-conducting alloy contacts (with coherence length $\xi \sim  550\text{nm}$) deposited on a monolayer graphene sheet. 
The scattered points are the data taken from the experimental work.
The behaviour of these plots can  be understood  by taking into account the variation of the superconducting gap $\Delta_{0}(T_{0})$
 with respect to the temperature $T_{0}$. The value for the gap is maximum when $T_{0} \rightarrow 0$ and minimum when $T_{0} \rightarrow T_{c}$. 
 For the above plots we have taken $T_{c}$ to be $5$K \cite{borzenets}, and as we can see the currents show a decrease with increase in temperature. The current approaches its maximum value as the temperature decreases. Qualitatively, the theoretical and experimental data match quite well.
 
Fig. \ref{JJexpt}(a)-(d)  show the variation of current as a function temperature for different gate voltages for various junction lengths $\emph{L}$. In Fig. \ref{JJexpt}(a)  for $\emph{L} = 200 \text{nm}$, which is  substantially less than the coherence length $\xi \sim  550 \text{nm}$, 
 the Josephson current varies between  $4 \mu A$ to $10 \mu A$ in the theoretical plots and from $2 \mu A$ to $7 \mu A$  in the experimental plots (as the gate voltage $\emph{V}$ varies in between $1.25 \text{V} - 13.5 \text{V}$.  However, in Fig. \ref{JJexpt}(d) for $\emph{L}= 2000nm$, the longest junction considered, the Josephson current now varies between $3 \mu A$ to $6.5 \mu A$ in theoretical plots and from $2.3 \mu A$ to $5 \mu A$ in the experimental plots
as the gate voltage  $\emph{V}$ is  varied from $3.98 \text{V} - 20.0 \text{V}$. Fig. \ref{JJexpt}(b) and (c) shows the behavior for intermediate junction lengths $\emph{L}= 400, 1000 \text{nm}$. 
Even though the theoretical calculations mostly overestimate the value of the current, which may be partially accounted for the  losses arising from contact resistance, non uniformity of the sample etc. which appears in a real junction and not taken into account in the calculation, following facts can be corroborated both from the theoretical and the experimental results. We see that the  maximum value of the Josephson current drops, as we go from short junctions to long junctions. The change of maximum Josephson current with the increase of the gate voltage also decreases as the $\emph{L}$ is increased.

In Fig. \ref{fig:ret_c2}(a), we plot the variation of the Josephson current with the gate voltage $V$ for several values of $T_{0}$ for a given length of the junction $\emph{L}=200 \text{nm}$. The gate voltage was taken on the both side of the Dirac point which lies close to $4.8\text{V}$ as was the case in the experiment \cite{borzenets}.
We see a dip around the Dirac point (close to 4.8V) and the current shows the bipolar behavior, both of which can be explained from the Eq.(\ref{universal}).\\
\begin{figure*}[!htb]
\subfloat[]{\includegraphics[width=.95\columnwidth]{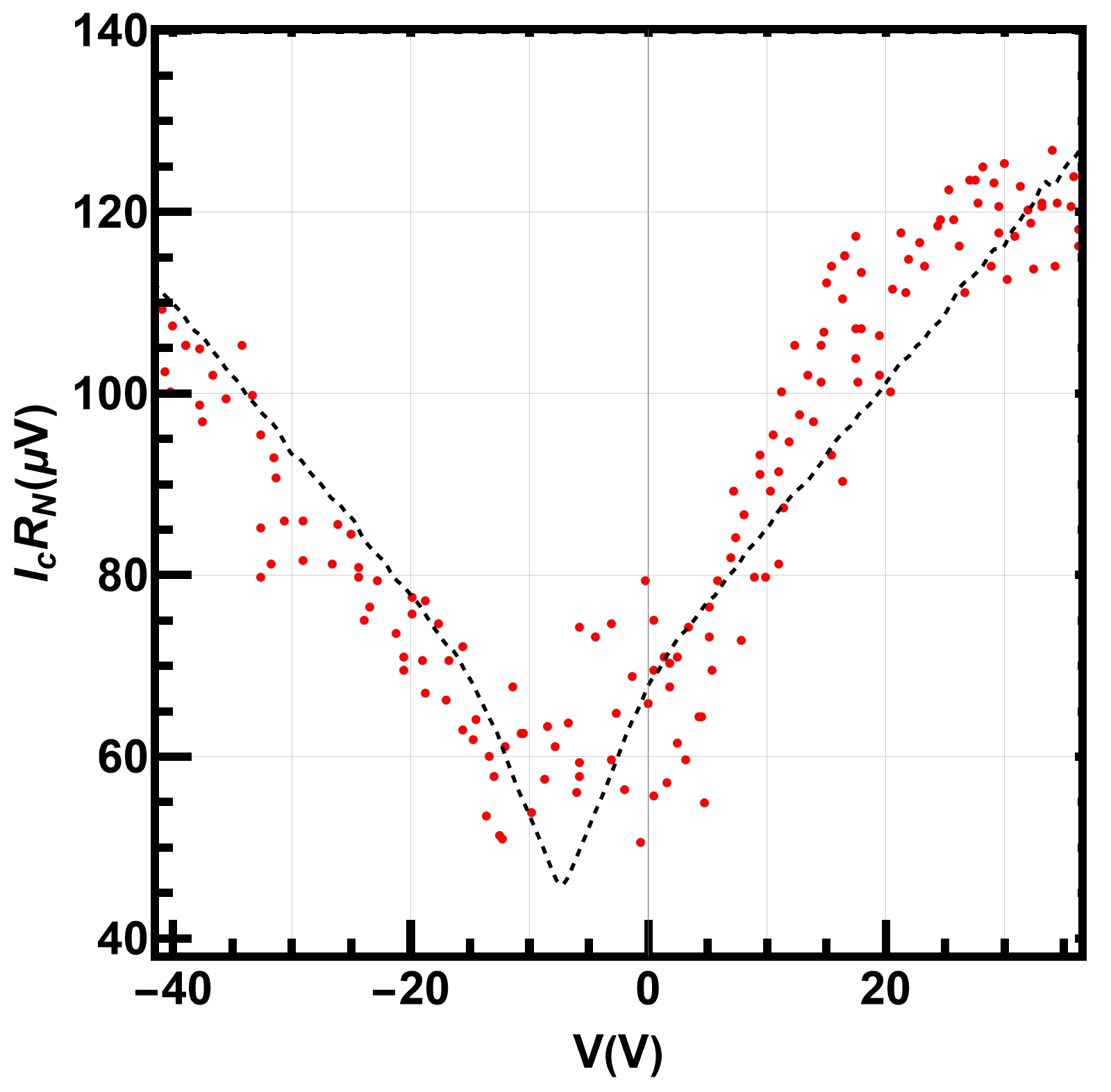}} 
\subfloat[]{\includegraphics[width=.95\columnwidth]{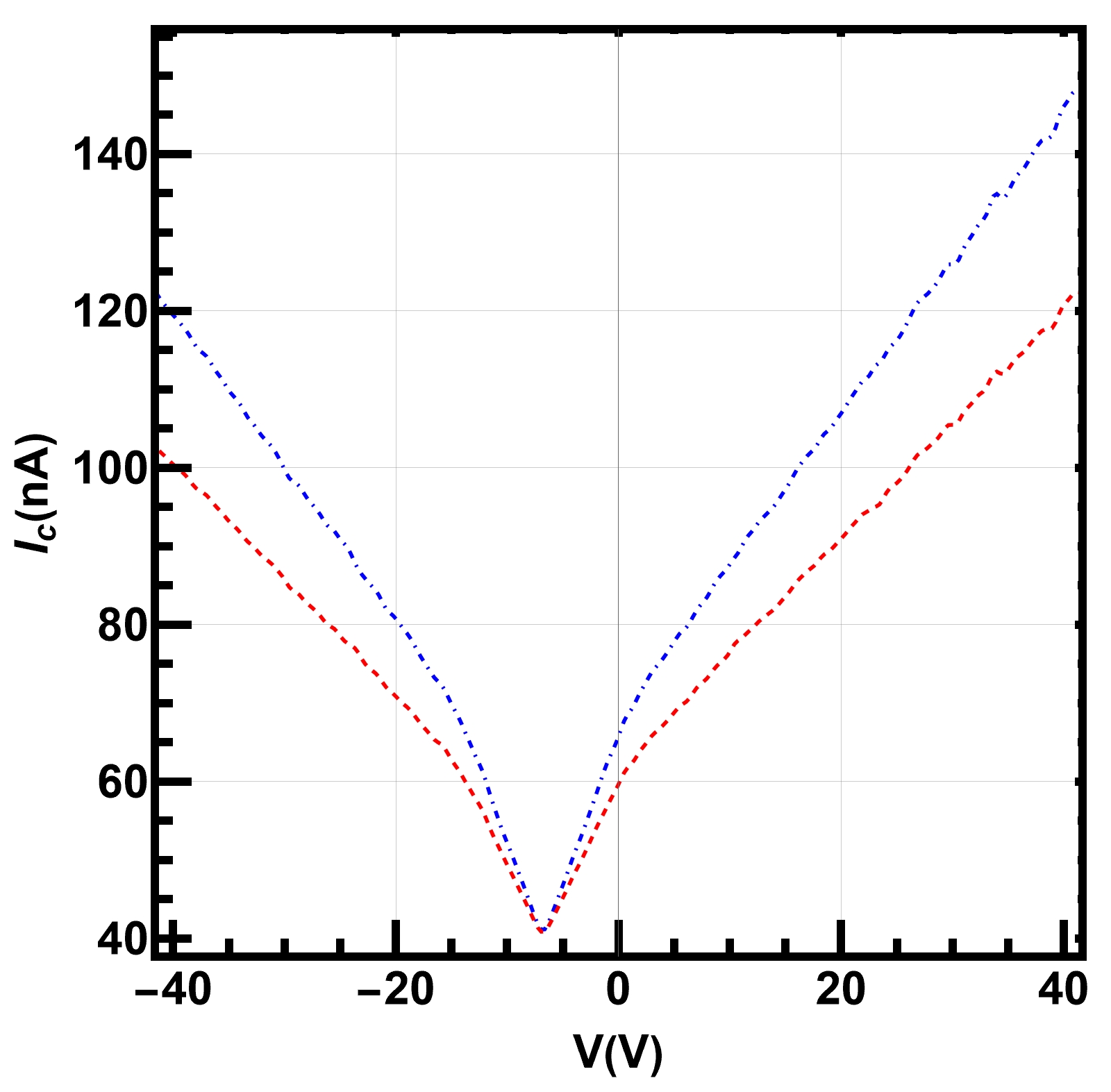}} 
 \caption{ (a) Comparison with the experimental value of $I_{c}R_{N}$ as a function of gate voltage. The scattered points are taken from the experimental results reported in \cite{Herrero}
whereas the dashed line is the theoretically calculated value. The superconducting gap is $\Delta_{0}(T_{0}=0)=125 \mu\text{eV}$
(b) $I_{c}$ as a function of the gate voltage at two different temperature, where $I_{c}$ is the critical Josephson current of the junction. The upper-plot corresponds to $T_{0}=1.0\text{K}$ and the lower plot corresponds to $T_{0}=1.25\text{K}$.}
\label{fig:ret_c}
\end{figure*}

Another direct comparison with the experimental results can be done by plotting the normal conductance $G_{N}$ (inverse of the normal state resistance $R_{N}$), as a function of the gate voltage $V$ from Eq. (\ref{conduct}) and this is plotted in 
 Fig. \ref{fig:ret_c2}(b)  and compared against the experimental observations \cite{borzenets} again for a  wide range of the length $L$ of the junction. 
 The normal state resistance $R_{N}$ is defined as 
\beq R_{N}^{-1} = \frac{\emph{LW}k_{F}}{8\pi} \frac{4e^{2}}{h} \int_{-k_{F}}^{k_{F}} t dq \label{conduct}, \eeq 
where the transmission coefficient $t$ is defined in Eq. (\ref{transmission}). This expression is similar to the one used in \cite{golubov, haberkorn}, and differs from the expression given in \cite{beenk_bal} by a constant factor. 
It may be noted that the transmission $t$ has been plotted as a function of the gate voltage $V$ in the inset  of the Fig.  \ref{fig:ret_tr}  on the both side of the Dirac points, and the integration limit $k_{F}$ is defined in the expression (\ref{params}).
The scaling of the conductance over a wide range of junction length is also clearly demonstrated through this comparison between the experimental and the theoretical results. 
 Both the experimental and theoretical plots show oscillatory features at large absolute values of the voltage $V$.  
 The nature of the plot can be understood from the fact that the transmission $t$ is a function of the incident angle $\alpha$, energy-gap ratio $\beta$ and other parameters which depend on the potential $V$. 
 When the applied potential is small, then $k_{e}\rightarrow k_{F}$ {\it i.e.} it becomes a constant. Furthermore the incident angle $\alpha \rightarrow 1$ when $V$ $\rightarrow 0$. Under these conditions the $\alpha$ dependence goes away. Thus behaviour in this  region is not oscillatory or the oscillations are very weak.  When the applied voltage is increased the oscillations start to appear and become more prominent. The theoretical plot based on the analytical expression given in Eq. (\ref{conduct}) also qualitatively captures the asymmetrical nature of the conductivity as a function of the gate voltage, even though quantitatively there is some discrepancy in the negative side of the gate voltage.

The bipolar behavior plotted in Fig. \ref{fig:ret_c2} (a), was also seen in a different experiment, reported in \cite{Herrero}. This is demonstrated in Fig. \ref{fig:ret_c}(a) and (b).
The experiment consists of superconducting $Ti/Al$ bilayers having maximum superconducting gap $\Delta_{0}(T_{0}=0)=125 \mu\text{eV}$ and the critical temperature $T_{c}=1.3\text{K}$. The critical current $I_{c}$ is calculated with the help of Eq. (\ref{JC1}) as well as the normal resistance $R_{N}$, are dependent on the gate voltage. The comparison of both $I_{c}$ and $I_{c} R_{N}$ between the experimental data from \cite{Herrero} and the theoretical results is done in Fig. \ref{fig:ret_c}. The agreement between the theoretically calculated value and the experimental result is again quite reasonable as can be seen from Fig. \ref{fig:ret_c}(a). It may be noted that 
superconducting gap here is  different ($125 \mu\text{eV}$) as compared to the value ($1.2 \text{meV}$) used in calculation designed for comparison with the experimental results reported in \cite{borzenets}. This also shows up in the magnitude and in the overall behavior of the critical current as a function of the gate voltage.
In Fig. \ref{fig:ret_c}(b) we plot the theoretically calculated  dependence of $I_{c}$ and $I_{c}R_{N}$ on the gate voltage $V$ for two 
 other temperatures. 

\subsection{Differential Resistance}\zlabel{rc}
The differential conductance can now be calculated by differentiating the expression for Josephson current  in Eq. (\ref{JC1}) with respect to the  bias voltage $U_{0}$. 
The resulting expression for differential conductance is given as 
\begin{widetext}
\begin{align}\label{eq:retcon}
&\frac{\partial}{\partial U_{0}} I(\phi,U_{0},V,T_{0})= I_{0}\int^{\frac{\pi}{2}} _{\frac{-\pi}{2}}  \frac{\partial}{\partial U_{0}} \left(
\frac{t \cos\gamma \sin\phi}{\sqrt{1-t\
sin ^{2}\frac{\phi}{2}}} 
\tanh\left(\frac{\Delta_{0}}{2k_{B}T_{0}}\sqrt{1-t\sin ^{2}\frac{\phi}{2}}- \frac{eU_{0}}{2k_{B}T_{0}}\right)   
   \right)d\gamma
\end{align}
\end{widetext}
~\\

\begin{figure}[!htb]
 \centering
\includegraphics[width=\columnwidth ]{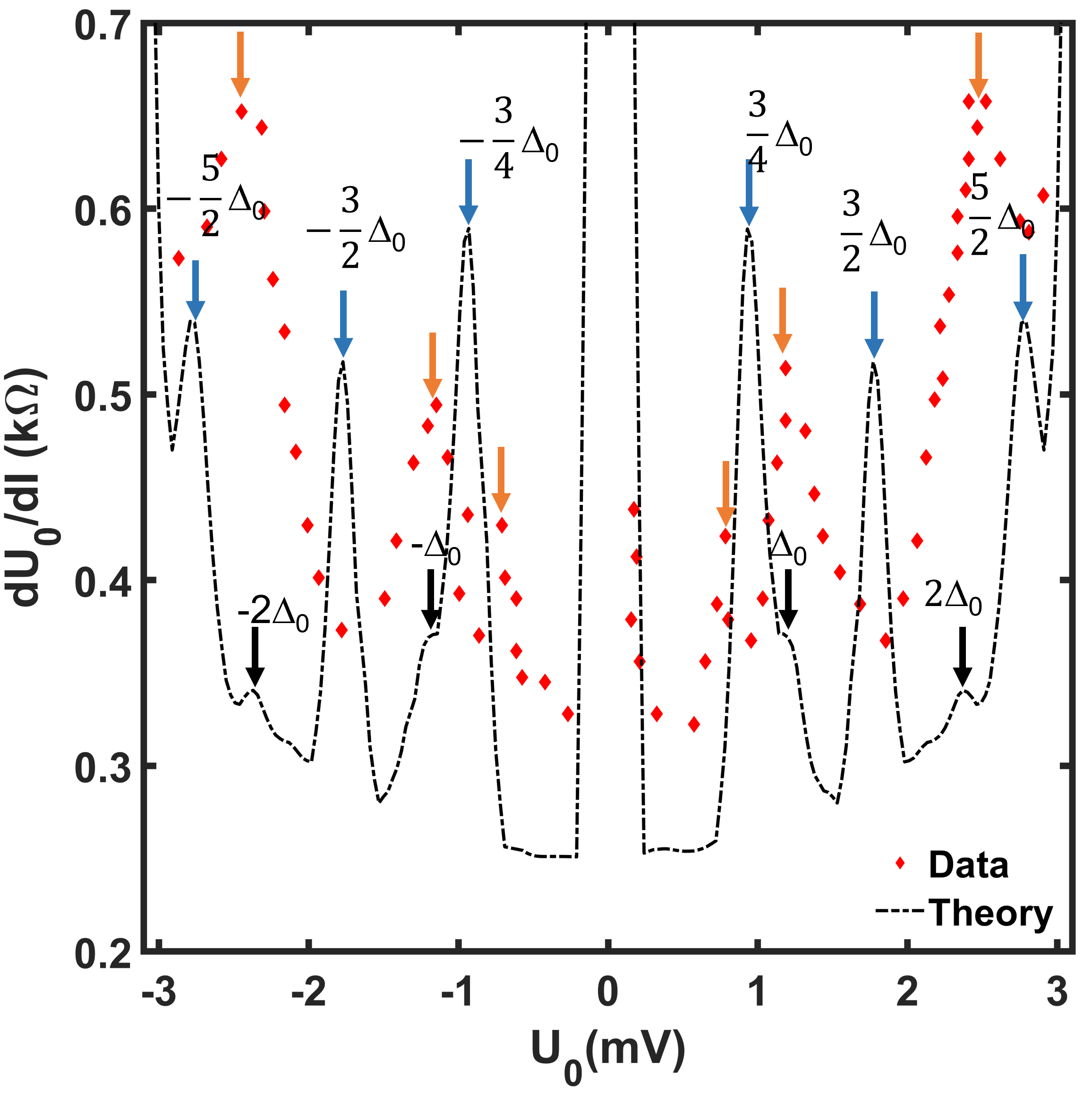}
\caption{Variation of the differential resistance as a function of the bias voltage $U_{0}$. The experimental data has been taken from Fig.1(b) of the experimental paper by Borzenets {\it et al.} \cite{borzenets}. Bias voltage applied is in milliVolts(mV). The differential resistance is of the order of $0.5-1~k\Omega$s. Several peaks are observed at regular intervals. }
\label{dvdi}
\end{figure}

The inverse of this expression gives the differential resistance, which is plotted in  Fig. \ref{dvdi} and compared with the experimental results\cite{borzenets}. It was observed that the differential resistance shows peaks for certain values of the biasing voltage that lie approximately around $\frac{2}{3}\frac{\Delta_{0}}{e}, \frac{\Delta_{0}}{e}, 2\frac{\Delta_{0}}{e}$, etc. The value of the differential resistance is of the order of $ \sim 0.5  k \Omega$. The theoretical and the experimental plots show good agreement with the position of some peaks, though the heights of peaks obtained from the theoretical results differ from the experimentally obtained results. We have used two different colours for the arrows to indicate the theoretically 
 obtained peaks at the integer and fractional multiples of $\frac{\Delta_{0}}{e}$ values of the biasing potential. Whereas we used another colour to 
indicate the positions of the peaks for the data points extracted from \cite{borzenets}. Such peaks occur because of the Fabry-Perot interferences  due to the reflections at the two SG interfaces, and, has been reported in a number of experiments done for SNS junctions \cite{FP1} as well as for SGS junctions \cite{FP2, FP3, FP4, FP5, FP6}.
The location of the peaks are different in different experiments. 
It may be pointed out that even though there is theoretical explanation \cite{KBT, OKBT} of the occurrence of such peak at $n \frac{\Delta_{0}}{e}$ where $n \in \mathcal{I}$ is available in the literature, a simple physical explanation of the occurrence of the other peaks is lacking at this moment. 
The comparison between the theoretically calculated resonance peak positions and the experimentally obtained one is another main result in this work. 

The limit of the highly doped superconducting Josephson junctions as shown in ref. \cite{beenk_sar},  by setting $\gamma \rightarrow 0$, $k \rightarrow U_0/\hbar v_F$, $\kappa \rightarrow(\Delta_0/ \hbar v_F)\sin\beta$, and correspondingly the expression of Josephson current will simplify. In the current manuscript we have given a simplified expression of transmission when $\gamma \rightarrow 0$ in expression (\ref{transmissionthin}).
 \begin{figure*}
 \centering
\subfloat[]{\includegraphics[width=0.5\columnwidth ]{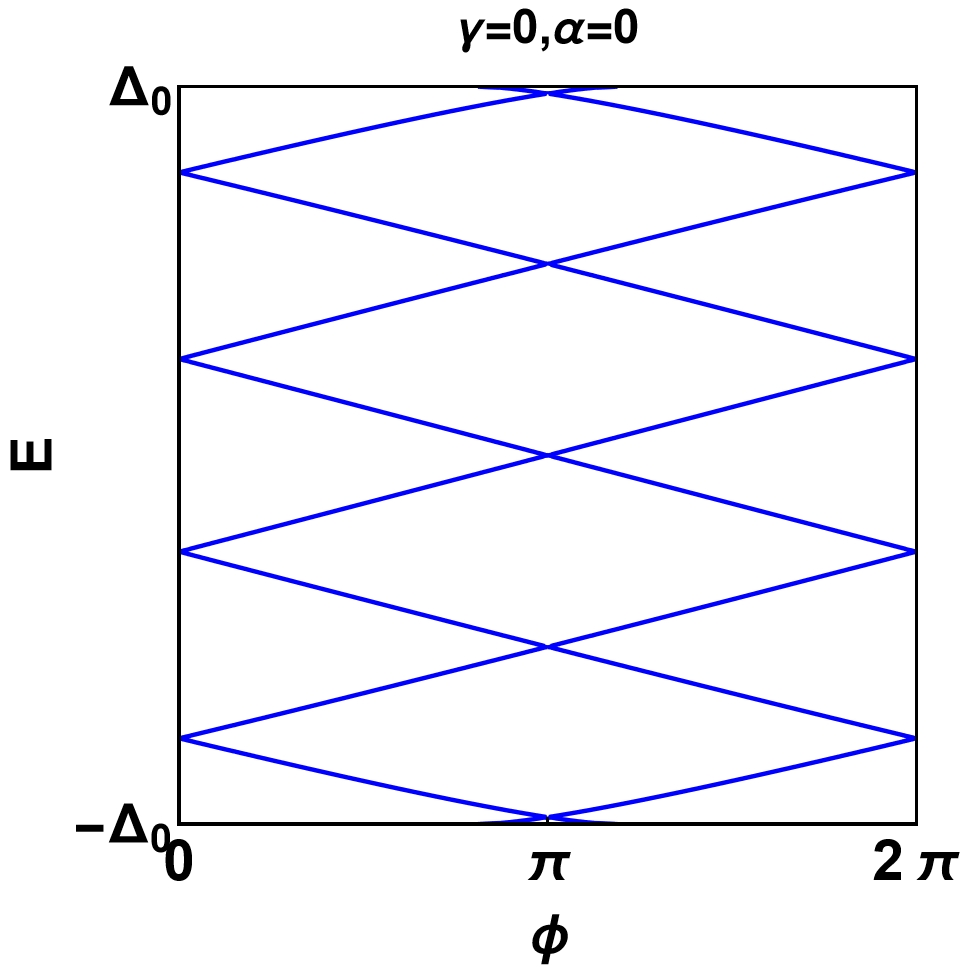}} 
\subfloat[]{\includegraphics[width=0.5\columnwidth]{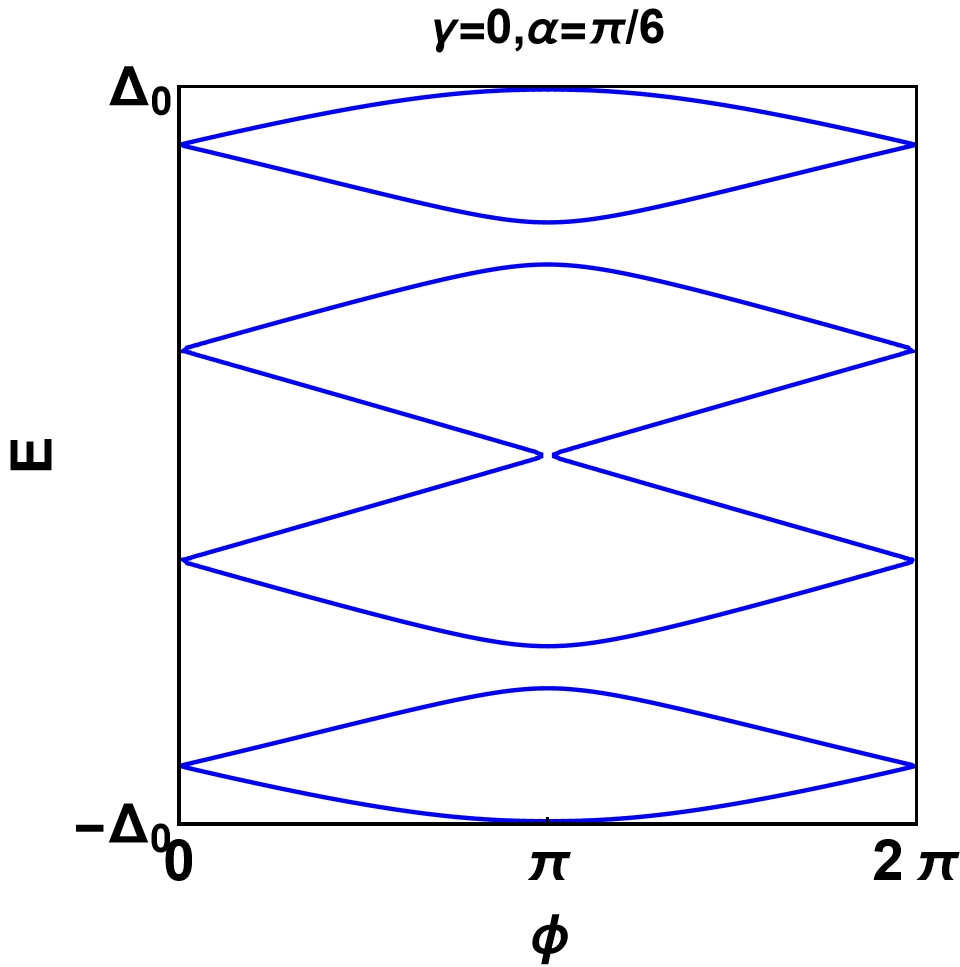}}
\subfloat[]{\includegraphics[width=0.5\columnwidth]{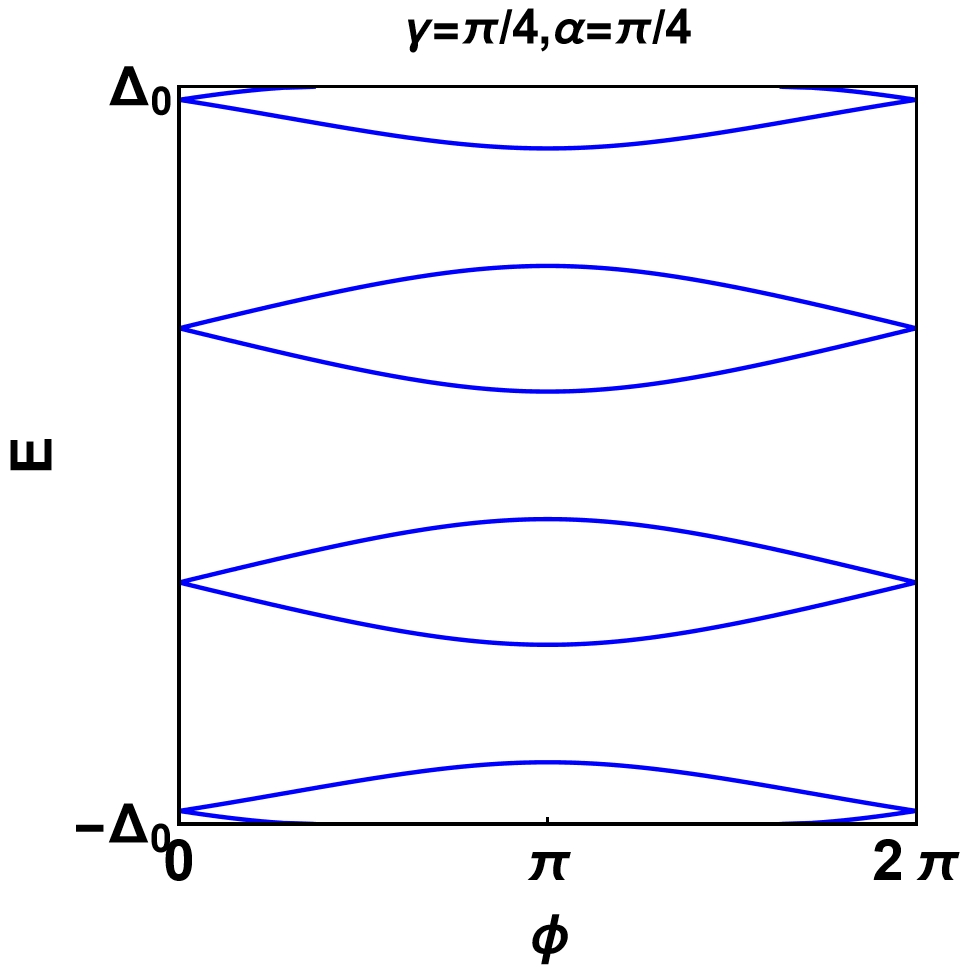}}
\subfloat[]{\includegraphics[width=0.5\columnwidth]{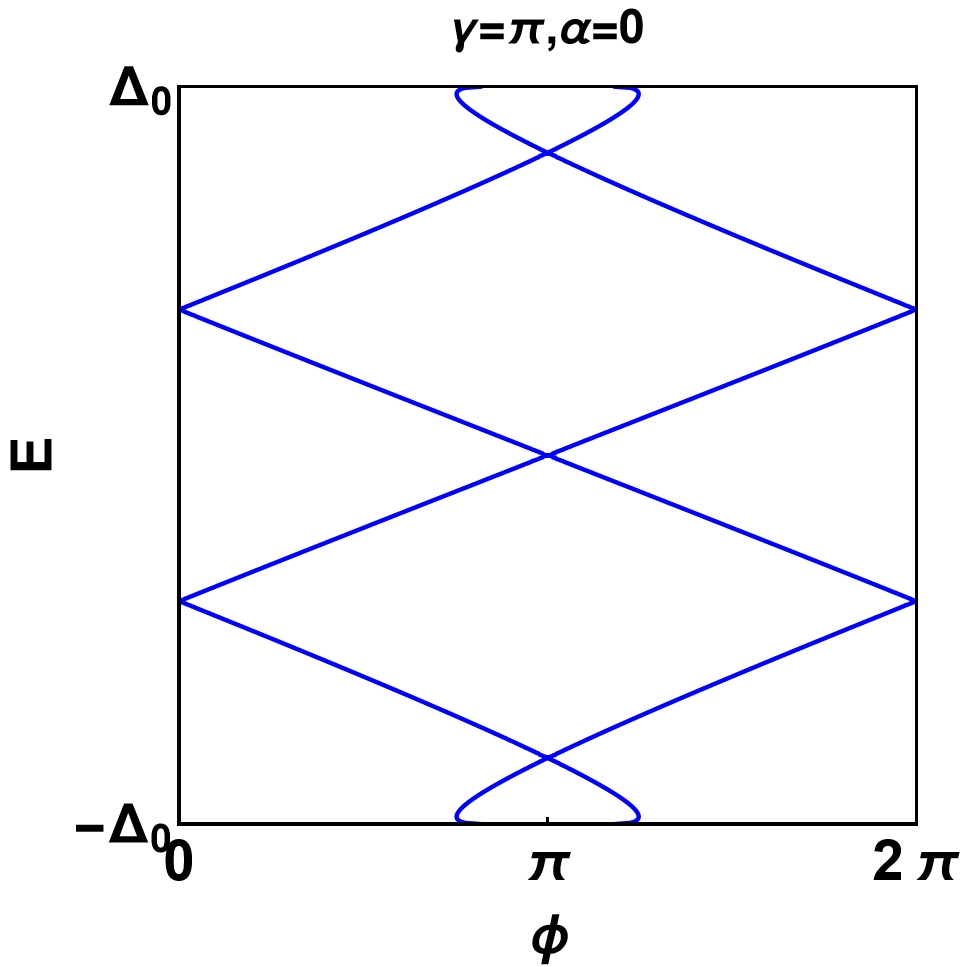}}\\
\subfloat[]{\includegraphics[width=0.52\columnwidth]{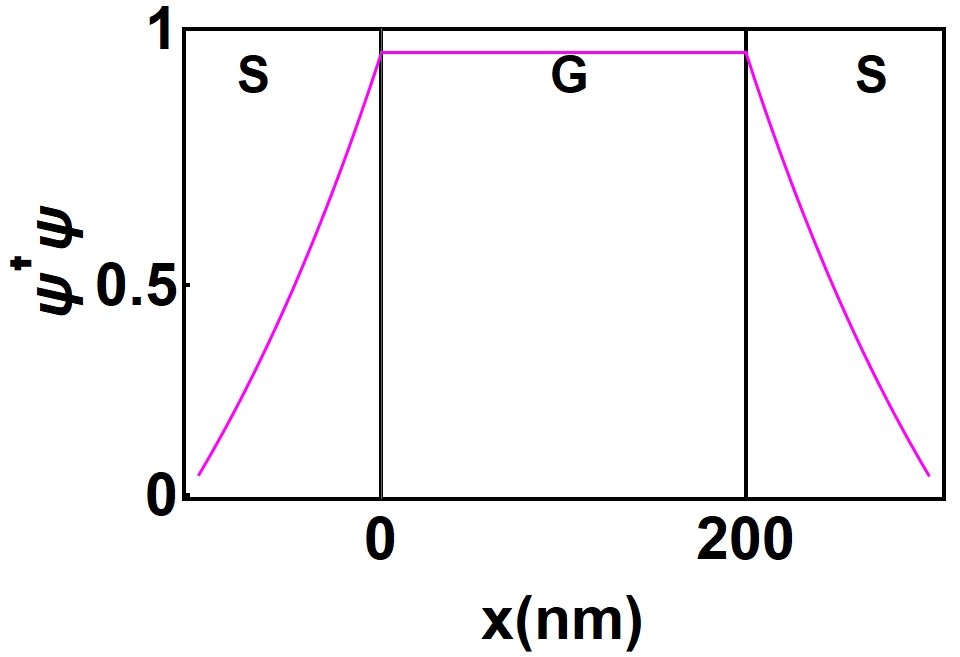}}
\subfloat[]{\includegraphics[width=0.47\columnwidth]{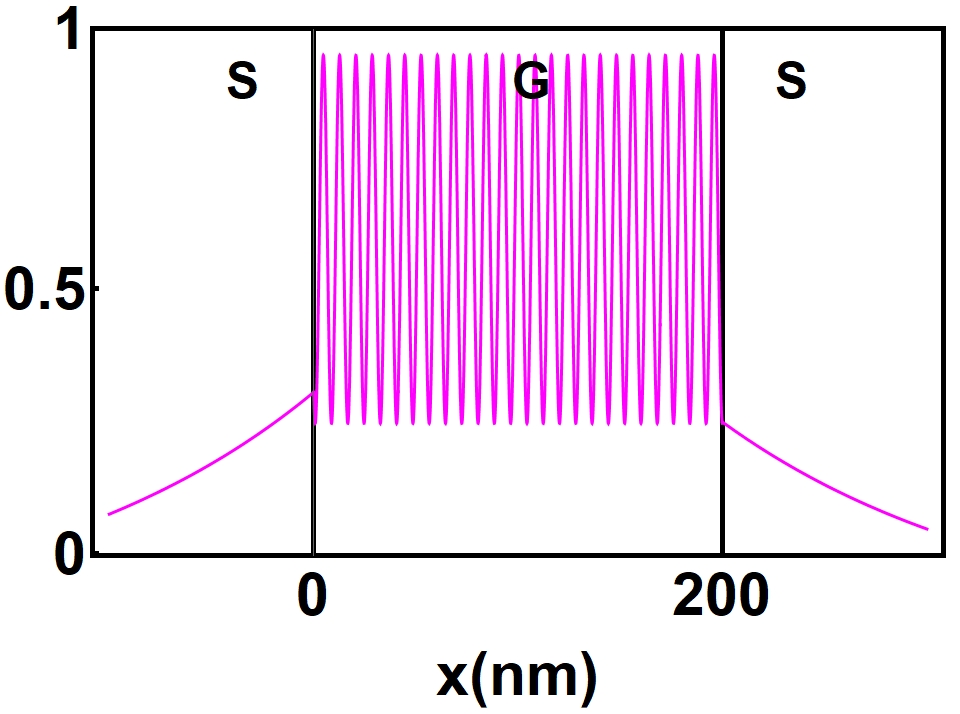}}
\subfloat[]{\includegraphics[width=0.47\columnwidth]{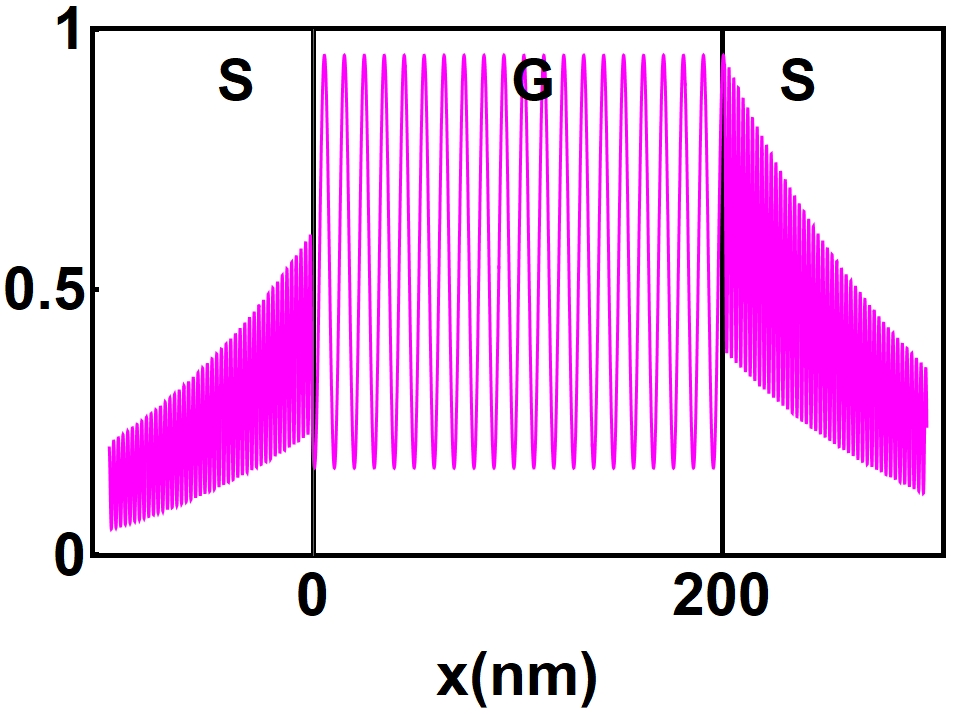}}
\subfloat[]{\includegraphics[width=0.47\columnwidth]{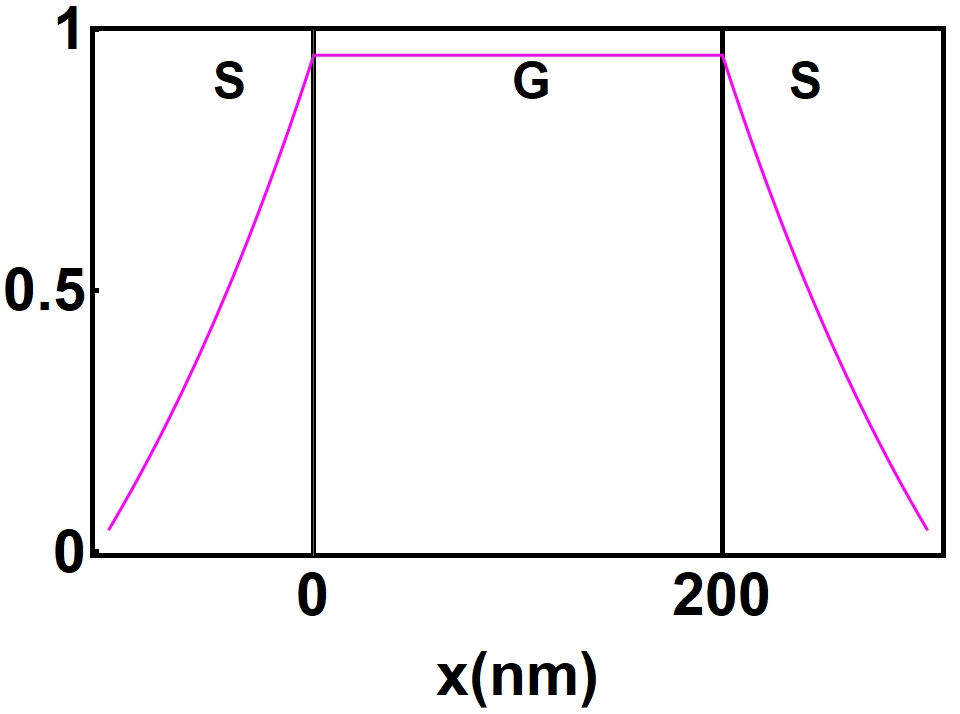}} 
\caption{(a-d) Dispersion for specular reflection. We see a large number of states for this case unlike the RAR case in Fig (\ref{retall1}). Maximum number of crossings are seen for the case when $\alpha=0,\gamma=0$. For all other values the gap widens and number of states reduce. (e-h) Prototype probability density corresponding to the respective dispersions in (a-d). The length $\emph{L}$ of the junction is $200$nm, $\phi_{1}= \frac{\pi}{4}$ and $\phi_{2}= \frac{\pi}{6}$ for all the figures.}
\label{specall}
\end{figure*}
\begin{figure*}[!htb]
\subfloat[]{\includegraphics[width =0.9\columnwidth]{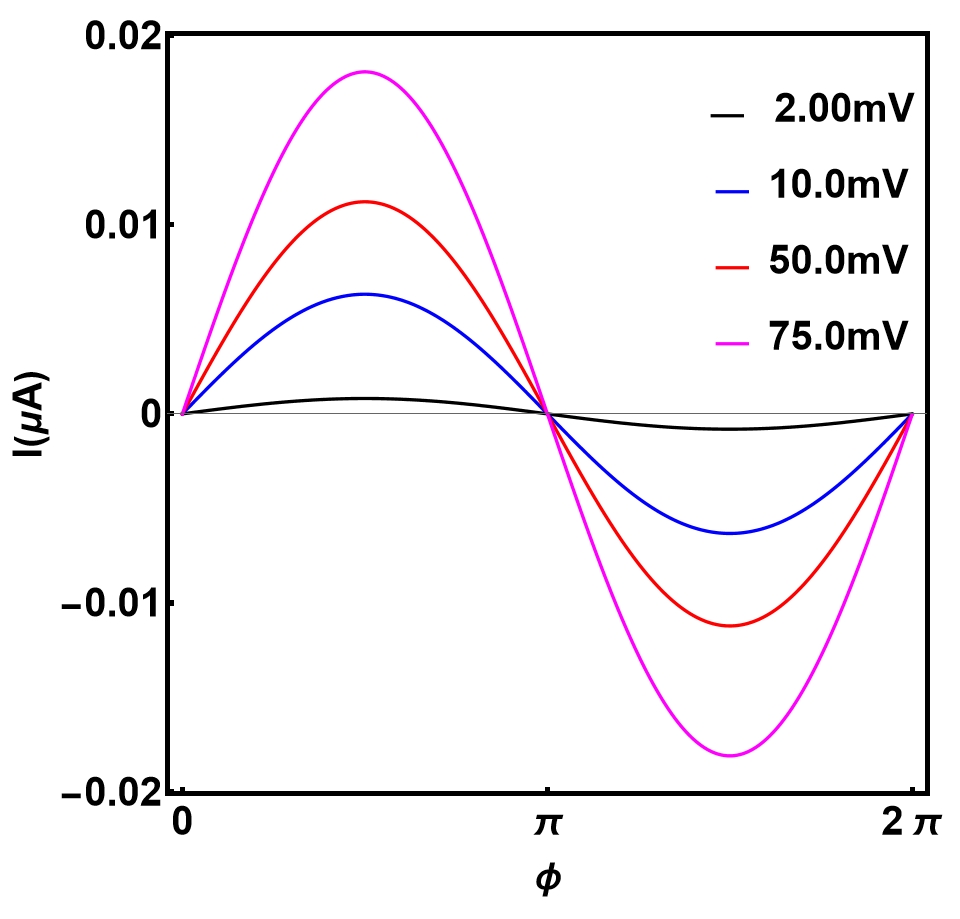}}
\subfloat[]{\includegraphics[width =0.9\columnwidth]{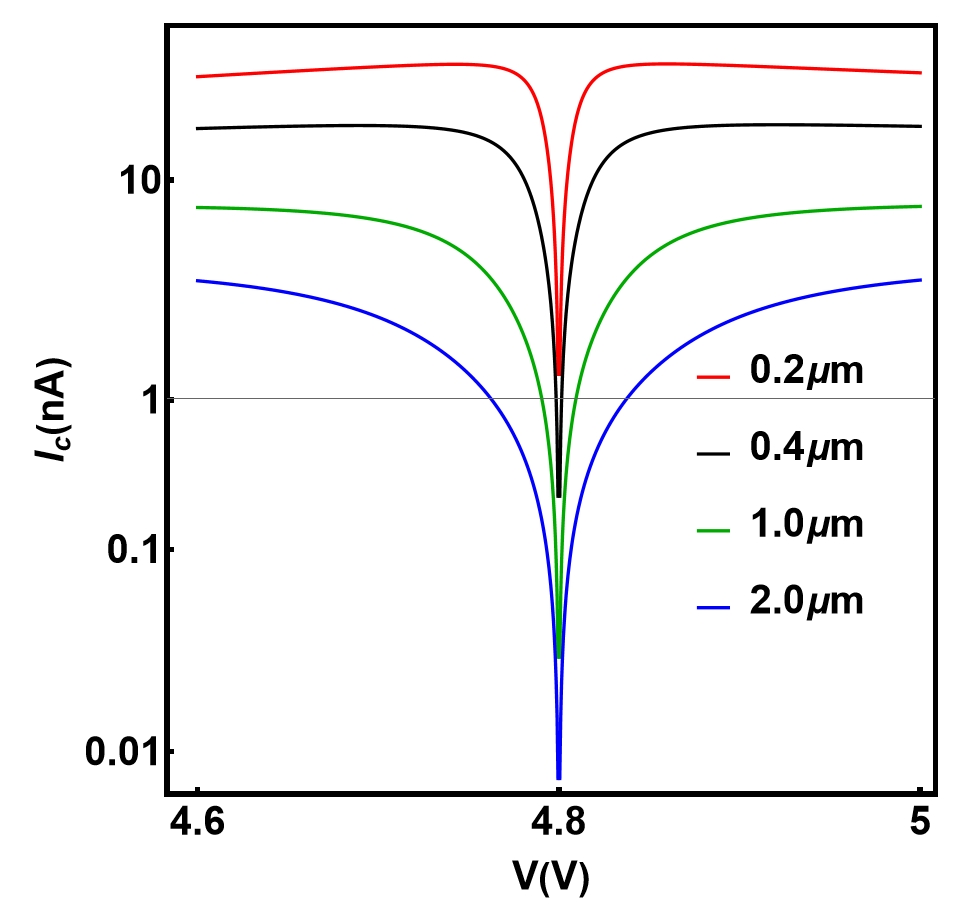}}
\caption{(a) Variation of the Josephson current $I$ due to SAR with the phase difference for different values of gate voltages. The temperature is kept at $2.5\text{K}$. (b) Variation of the critical Josephson current $I_{c}$ due to SAR as a function of the gate voltage for different lengths of 
the junction. The Dirac point is again kept at $V=4.8$ volt as in the experiment reported in \cite{borzenets}. The temperature  is same as in part(a). }
\label{specjosf}
\end{figure*}
\section{Specular Reflection}\zlabel{spar}
Specular Andreev reflection takes place when the incident and the reflected particle lie in different bands i.e. an electron in conduction band is converted into a hole in the valence band \cite{pothier,timm}. 
This happens when $E \ge E_{F}^{eff}$ and as a result we get
\beq \alpha '  =  \alpha;~k_{h} = k_{e}; \nonumber \eeq 
The general specular dispersion relation can again be given by the Eq. (\ref{conic}) with $A, B, C, D$ given in Eqs. (\ref{disparam}). As compared to retro reflection $C \neq 0$ in this case.
The corresponding  dispersion has been plotted in Fig. \ref{specall}. 
The dispersion equation can be reduced to a simpler form as $\alpha = \gamma = 0$,  giving 
 \beq \sin^{2}\left( \beta +\frac{LE}{\hbar v_{F}}\right) = \sin ^{2}\frac{\phi}{2}. \nonumber \eeq 
 The corresponding results are plotted in Fig. \ref{specall}(a) yielding a linear dependence between $E$ and $\phi$.
The general form of the dispersion  given in Eq. (\ref{conic}) for the specular case is also valid for both short and long junctions and have many solutions in the allowed energy range \cite{sauls, Bretheau}. Fig. \ref{specall}(a)-(d) 
give the dispersion as a function of the superconducting phase difference $\phi$. Fig. \ref{specall}(e)-(h)  plot the probability densities for some prototype ABS for SAR. \\

A special case occurs when $V \rightarrow E_{F}$, namely when $E_{F}^{eff} \rightarrow 0$. To address this situation we go back to the original definition of these angles in Eq. (\ref{eq:incident}). In the limit $V\rightarrow E_{F}$, one can redefine these angles as
\begin{eqnarray}\label{eq:newinc}
\sin\alpha = \frac{\hbar v_{F}q}{E},~ 
\sin\alpha ' = \frac{\hbar v_{F}q}{E}
\end{eqnarray}
which makes the incident angle for electron and hole equal. This situation  also corresponds to that of specular reflection. So, at $V=E_{F}$; $\alpha = \alpha '$, we get Specular Andreev Reflection (SAR). Exactly ascertaining such transition in an experiment is however difficult because of the strong charge fluctuations at the Fermi-level \cite{Yacoby, anindya}. In this case the expression of the Josephson current also becomes the one in the case of the specular Andreev reflection.

\subsection{ Josephson Current and the differential resistance}
\zlabel{sjc}
The Josephson current can be obtained from the  dispersion relation by noting 
\beq 
\frac{\partial \phi}{\partial E} = \phi_{c}' \cos Lk_{e} + \phi_{s}'\sin Lk_{e}
 \eeq 
where 
\begin{widetext}
\bea 
\phi_{c}' & = & \frac  {-E\sqrt{1- (\frac{\hbar v_{F}q}{E})^{2}}\left(\Delta_{0} L E^{2}\sqrt{1-\frac{E^{2}}{\Delta_{0}^{2}}} +E^{2}\hbar v_{F} \sqrt{1- (\frac{\hbar v_{F}q}{E})^{2}} \sec\gamma\right)}{ \hbar v_{F}E^{2}\Delta_{0}^{2}\sin\beta\cos ^{3}\alpha} \nonumber \\
\phi_{s}' & = & \frac{ \left(-\hbar v_{F}\Delta_{0} \sqrt{1- \frac{E^{2}}{\Delta_{0}^{2}}}(E^{2} - 2(\hbar v_{F}q)^{2}) +L(E^{2}-\Delta_{0}^{2})(E^{2} - (\hbar v_{F}q)^{2})\sec\gamma\right)}{ \hbar v_{F}E^{2}\Delta_{0}^{2}\sin\beta\cos ^{3}\alpha} \nonumber 
\eea 
\end{widetext}

The Josephson current (\ref{jcur1}) can now be calculated as
\beq
I(\phi;U_{0},T_{0})=\frac{4e}{\hbar}\sum_{n} \sum_{q=-k_{F}} ^{k_{F}} \big{[}\frac{\partial \phi}{\partial E} \big{]}^{-1}_{(E=E_{n})} 
\tilde{f}(E_{n}) \label{JJCSAR}
\eeq 
 Fig.\ref{specjosf}(a) plots the current given through the expression (\ref{JJCSAR}) as a function of the superconducting phase-difference for different values of the gate voltage. As expected that the current phase relationship for SAR is similar to the one depicted in Fig. \ref{JCFig}, however the relative magnitude of the current is 
 much smaller as compared to the one due to RAR, thus makes them unobservable in the current experiments.
 In Fig. \ref{specjosf}(b) we plot the corresponding critical Josephson current as a function of the gate voltage $\emph{V}$ , for a fixed $T_{0}=2.5\text{K}$, for different lengths $\emph{L}$ of the junction. The current again shows the bipolar behaviour around the Dirac point which is located 
 at $V=4.8~\text{volt}$ to make a direct comparison with the results given in Fig. \ref{fig:ret_c2}.
\begin{figure*}
 \centering
\subfloat[]{\includegraphics[width=0.95\columnwidth]{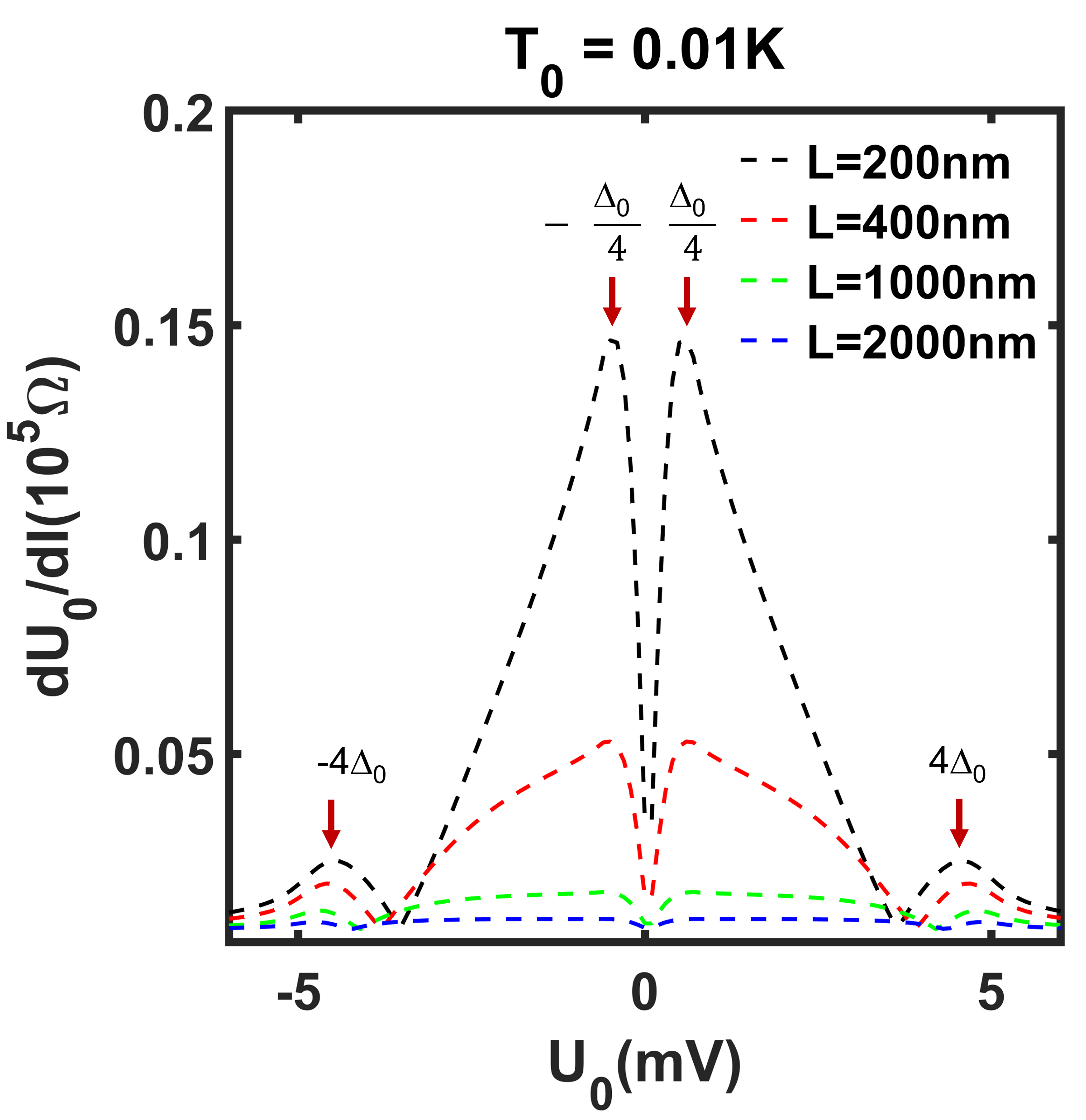}}
\subfloat[]{\includegraphics[width=0.95\columnwidth]{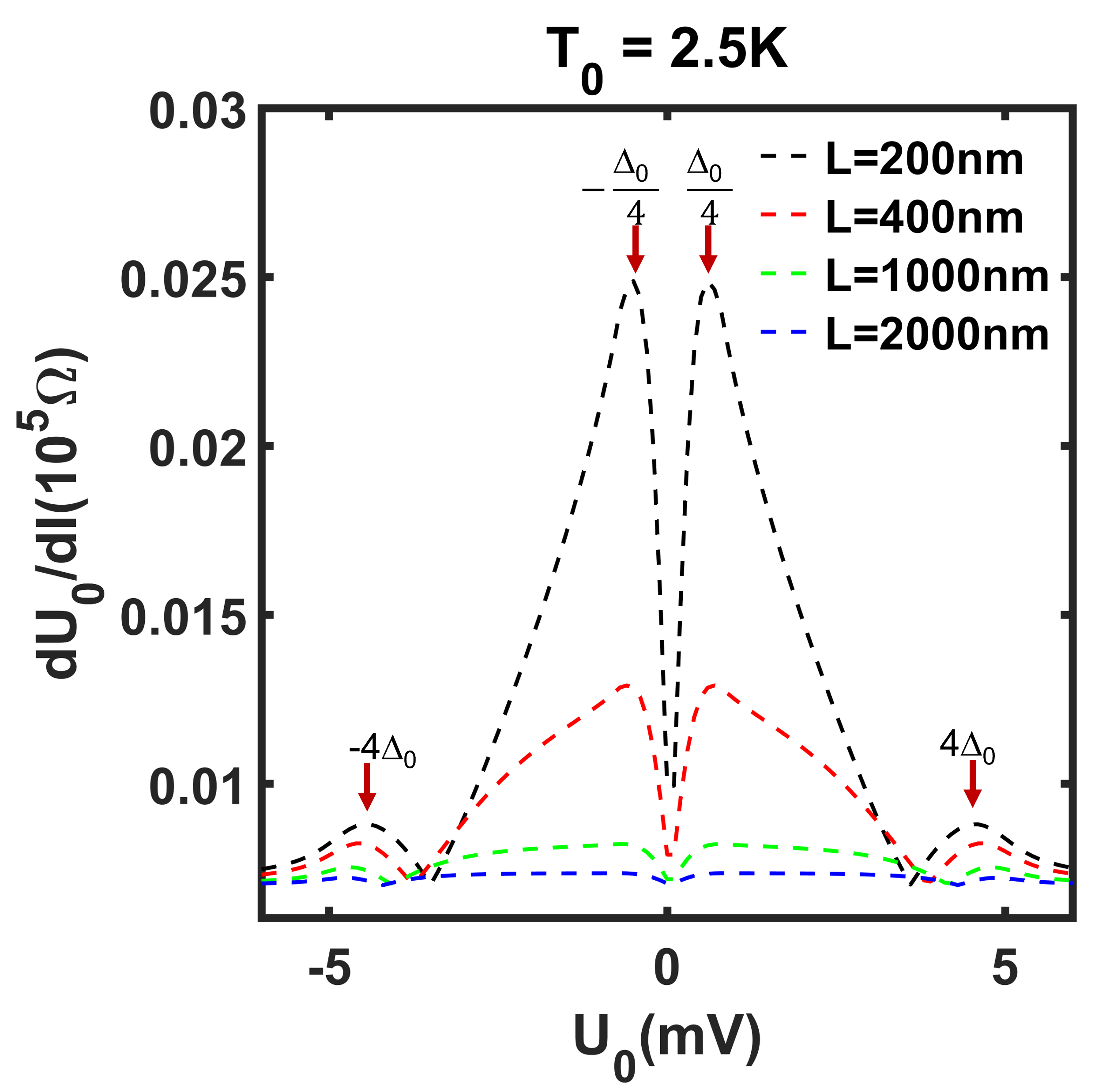}}
\caption{Variation of the differential resistance
with  the bias voltage $U_{0}$ for different junction  lengths $\emph{L}=200, 400, 1000, 2000 \text{nm}$ for (a) $T_{0}=0.01$ and (b) $T_{0}=2.5K$. The location of the resistance peaks are again denoted by coloured arrows in the same way as in the Fig. \ref{dvdi}.}
\label{spcall}
\end{figure*}
 
The differential conductance can again be calculated by taking the derivative of the above expression with respect to bias voltage $U_{0}$ and the differential resistance can be obtained by taking the inverse of the same.
  In Fig. \ref{spcall} we plot the differential resistance for different length of the junctions at two different values of the temperature $T_{0}$. We see two sharp peaks at $U_{0} = \pm\frac{\Delta_{0}}{4e}$ and two smaller peaks at $U_{0} = \pm 4 \frac{\Delta_{0}}{e}$. These peaks can again be attributed to Fabry-Perrot resonances, but they are now formed due to the current carried by Andreev bound states due to SAR. The heights of the peaks are largest for the smallest length of the junctions considered.
 The values of the differential resistances are typically an order of magnitude higher as compared to the corresponding cases for RAR plotted in Fig. \ref{dvdi} because of the relatively smaller value of the critical current.
~\\~\\~\\
\section{Conclusion}
\zlabel{conc}
In conclusion we have provided a detailed comparison between the experimentally observed currents and conductances with the theoretically calculated results for a SGS type of Josephson junction. 
The theoretical calculation is done using a transfer-matrix based approach and for all the relevant quantities we have obtained analytical expressions. The formation of the ABS for both RAR and SAR processes were thoroughly investigated in this framework and their energy values were then used to calculate the Josephson current and the conductance. 
The critical currents calculated for various junction-lengths and at various temperatures agree qualitatively well 
with the experimentally obtained results. The direct comparison with the experimentally obtained results also  provide a quantitative estimate of the deviations from the experimentally calculated quantities and may provide a a better theoretical understanding in future. 
In this context it may be pointed out that the massless Dirac approximation for monolayer graphene is only valid upto typically $\sim 3 \text{eV}$ from the Dirac point, beyond which higher order terms from the tight-binding calculation ( for example see ref. \cite{Diracapprox}) becomes important. 
It may be a noted that in a set of figures where we have tried to compare our theoretical results with experimental data, the results are extended beyond this range whereas the DBdG Hamiltonian given in Eq. (\ref{eq:DBdG}) is based on the Dirac approximation. So one possibility of the quantitative deviation between the theoretical calculation and the experimental results may be partially attributed to the the deviation of the Dirac approximation itself. We hope future work in this direction will address this issue. Furthermore, it may be noted that a different choice of Fermi energy will quantitatively change a number of results and such issues were considered in earlier works ( e. g. see ref. \cite{Fermichange}).
Similarly the analytically calculated conductance as a function of gate voltage also shows strong scaling properties with all its features agreeing well with the experimental results. We also evaluated the differential resistance as a function of the bias voltage and compared with the experimental results. Locations of 
some of the resonance peaks in the differential resistance agree well with the locations of such peaks obtained in the experiments. We have also evaluated, the current-phase relationship, the variations of the critical current with the gate voltage and the differential resistivity when SAR dominates. Whereas the current obtained in this  case is significantly smaller than their counterparts in the case of RAR, the differential resistance as a function of the bias voltage is significantly higher and also formed distinct resonance peaks. We believe that our detailed analytical work and comparison with experiments can lead to further experimental and theoretical work.

\appendix
\section{The detailed expression for the transfer matrices}
\label{app1}
\begin{widetext}
\bea 
M_{1} & = & \begin{bmatrix}
e^{i\beta}& e^{i\beta} & e^{-i\beta} &e^{-i\beta}\\
e^{i\beta}& -e^{i\beta} & -e^{-i\beta} &e^{-i\beta}\\
e^{-i\phi_{1}} & e^{-i\phi_{1}} & e^{-i\phi_{1}} & e^{-i\phi_{1}}\\
e^{-i\phi_{1}} & -e^{-i\phi_{1}} & -e^{-i\phi_{1}} & e^{-i\phi_{1}}
\end{bmatrix}  \nonumber \\
M_{2} & = & \begin{bmatrix}
\frac{e^{-\frac{i\alpha}{2}}}{\sqrt{\cos\alpha}}& \frac{e^{\frac{i\alpha}{2}}}{\sqrt{\cos\alpha}}&0&0\\
\frac{e^{\frac{i\alpha}{2}}}{\sqrt{\cos\alpha}}& -\frac{e^{-\frac{i\alpha}{2}}}{\sqrt{\cos\alpha}}&0&0\\
0&0& \frac{e^{-\frac{i\alpha '}{2}}}{\sqrt{\cos\alpha '}}& \frac{e^{\frac{i\alpha '}{2}}}{\sqrt{\cos\alpha '}}\\
0&0& -\frac{e^{\frac{i\alpha '}{2}}}{\sqrt{\cos\alpha '}}& \frac{e^{-\frac{i\alpha '}{2}}}{\sqrt{\cos\alpha '}}
\end{bmatrix} \nonumber \\
M_{3} & = &  \begin{bmatrix}
\frac{e^{ik_{e}L-i\frac{\alpha}{2}}}{\sqrt{\cos\alpha}}& \frac{e^{-ik_{e}L+i\frac{\alpha}{2}}}{\sqrt{\cos\alpha}}& 0&0\\
\frac{e^{ik_{e}L+i\frac{\alpha}{2}}}{\sqrt{\cos\alpha}}& -\frac{e^{-ik_{e}L-i\frac{\alpha}{2}}}{\sqrt{\cos\alpha}}& 0&0\\
0&0&\frac{e^{ik_{h}L-i\frac{\alpha '}{2}}}{\sqrt{\cos\alpha '}}& \frac{e^{ik_{h}L+i\frac{\alpha '}{2}}}{\sqrt{\cos\alpha '}}\\
0&0&-\frac{e^{ik_{h}L+i\frac{\alpha '}{2}}}{\sqrt{\cos\alpha '}}& \frac{e^{-ik_{h}L-i\frac{\alpha '}{2}}}{\sqrt{\cos\alpha '}} 
\end{bmatrix} \nonumber \\
M_{4} & = & \begin{bmatrix}
e^{ik_{0}L-\kappa L+ i\beta}&e^{-ik_{0}L+\kappa L+i\beta}&e^{-ik_{0}L-\kappa L-i\beta}& e^{ik_{0}L+\kappa L -i\beta}\\
e^{ik_{0}L-\kappa L+ i\beta}& -e^{-ik_{0}L+\kappa L+i\beta}& -e^{-ik_{0}L-\kappa L-i\beta}& e^{ik_{0}L+\kappa L -i\beta}\\
e^{ik_{0}L-\kappa L- i\phi_{2}}&e^{-ik_{0}L+\kappa L- i\phi_{2}}&e^{-ik_{0}L-\kappa L- i\phi_{2}}& e^{ik_{0}L+\kappa L - i\phi_{2}}\\
e^{ik_{0}L-\kappa L- i\phi_{2}}&e^{-ik_{0}L+\kappa L- i\phi_{2}}&e^{-ik_{0}L-\kappa L- i\phi_{2}}& e^{ik_{0}L+\kappa L - i\phi_{2}}
\end{bmatrix} \label{transfer}
\eea 
\end{widetext}

\end{document}